\documentclass[zpreprint,zbstnp]{zeus_paper}

\usepackage[english]{babel}

\newcommand{\Zpsrap}{%
The pseudorapidity is defined as $\eta=-\ln\left(\tan\frac{\theta}{2}\right)$,
where the polar angle, $\theta$, is measured with respect to the proton beam
direction.\xspace}

\newcommand{\Zdetdesc}{%
A detailed description of the ZEUS detector can be found 
elsewhere~\cite{zeus:1993:bluebook}. A brief outline of the 
components most relevant for this analysis is given
below.\xspace}
\newcommand{\Zctddesc}[1]{%
Charged particles were tracked in the central tracking detector (CTD)~\citeCTD,
which operated in a magnetic field of $1.43\Tesla$ provided by a thin 
superconducting coil. The CTD consisted of 72~cylindrical drift chamber 
layers, organized in 9~superlayers covering the polar-angle#1 region 
\mbox{$15^\circ<\theta<164^\circ$}. }

\newcommand{\Zcaldesc}{%
The high-resolution uranium--scintillator calorimeter (CAL)~\citeCAL consisted 
of three parts: the forward (FCAL), the barrel (BCAL) and the rear (RCAL)
calorimeters. Each part was subdivided transversely into towers and
longitudinally into one electromagnetic section (EMC) and either one (in RCAL)
or two (in BCAL and FCAL) hadronic sections (HAC). The smallest subdivision of
the calorimeter is called a cell.  The CAL energy resolutions, as measured under
test-beam conditions, are $\sigma(E)/E=0.18/\sqrt{E}$ for electrons and
$\sigma(E)/E=0.35/\sqrt{E}$ for hadrons ($E$ in $\Gev$).}

\chardef\usc=95
\chardef\til=126
\catcode`\@=11 
\DeclareRobustCommand\xdotspace{\futurelet\@let@token\@xdotspace}
\def\@xdotspace{%
  \ifx\@let@token.\else
  \ifx\@let@token\bgroup.\else
  \ifx\@let@token\egroup.\else
  \ifx\@let@token\/.\else
  \ifx\@let@token\ .\else
  \ifx\@let@token~.\else
  \ifx\@let@token!.\else
  \ifx\@let@token,.\else
  \ifx\@let@token:.\else
  \ifx\@let@token;.\else
  \ifx\@let@token?.\else
  \ifx\@let@token/.\else
  \ifx\@let@token'.\else
  \ifx\@let@token).\else
  \ifx\@let@token-.\else
  \ifx\@let@token\@xobeysp.\else
  \ifx\@let@token\space.\else
  \ifx\@let@token\@sptoken.\else
   .\space
   \fi\fi\fi\fi\fi\fi\fi\fi\fi\fi\fi\fi\fi\fi\fi\fi\fi\fi}
\catcode`\@=12 

\newcommand{\stru}[2]{%
   \relax\ifmmode\hbox{\vrule height#1 depth#2 width0pt}%
   \else\vrule height#1 depth#2 width0pt\fi}

\newcommand{\Ronum}[1]{\uppercase\expandafter{\romannumeral#1}}
\newcommand{\ronum}[1]{\expandafter{\romannumeral#1}}
\DeclareRobustCommand{\LaTeXZ}{%
  \LaTeX\kern-.05em4\kern-.1em
  {\raisebox{-0.2ex}{$\scriptstyle\text{ZEUS}$}}\xspace}

\newcommand{\eq}[1]{(\ref{eq-#1})}

\newcommand{\fig}[1]{Fig.~\ref{fig-#1}}
\newcommand{\Fig}[1]{Figure~\ref{fig-#1}}
\newcommand{\figand}[2]{Figs.~\ref{fig-#1} and~\ref{fig-#2}}

\newcommand{\tab}[1]{Table~\ref{tab-#1}}

\newcommand{\Sect}[1]{Section~\ref{sec-#1}}

\DeclareMathAlphabet{\mathbf}{OT1}{cmr}{bx}{sl}
\newcommand{\eVdist}{\kern-0.06667em}

\newcommand{\Gev}{{\text{Ge}\eVdist\text{V\/}}}

\newcommand{\gev}{{\,\text{Ge}\eVdist\text{V\/}}}

\newcommand{\nb}{\,\text{nb}}
\newcommand{\pb}{\,\text{pb}}

\newcommand{\cm}{\,\text{cm}}

\newcommand{\mrad}{\,\text{mrad}}

\newcommand{\Tesla}{\,\text{T}}

\newcommand{\slashfrac}[2]{%
  \raisebox{0.5ex}{\ensuremath #1}\kern-0.12em/\kern-0.08em
  \raisebox{-.8ex}{\ensuremath #2}}

\newcommand{\sqr}[3]{%
    {\vcenter{\hrule height.#3ex\hbox{\vrule width.#2ex height#1ex
     \kern#1ex\vrule width.#3ex}\hrule height.#2ex}}}

\catcode`\@=11 
\newcommand{\parenbar}{\mathpalette\p@renb@r}
\def\p@renb@r#1#2{\vbox{%
  \ifx#1\scriptscriptstyle \dimen@.7em\dimen@ii.2em\else
  \ifx#1\scriptstyle \dimen@.8em\dimen@ii.25em\else
  \dimen@1em\dimen@ii.4em\fi\fi \offinterlineskip
  \ialign{\hfill##\hfill\cr
    \vbox{\hrule width\dimen@ii}\cr
    \noalign{\vskip-.3ex}%
    \hbox to\dimen@{$\mathchar300\hfil\mathchar301$}\cr
    \noalign{\vskip-.3ex}%
    $#1#2$\cr}}}
\catcode`\@=12 

\newcommand{\IP}{{\rm I$\kern-0.01667em$P}\xspace}

\mathchardef\qsm=63
\mathchardef\pls=43
\mathchardef\mns=512
\mathchardef\plm=518
\mathchardef\eql=61
\mathchardef\smallleft=300
\mathchardef\smallright=301
\mathchardef\les=316
\mathchardef\gre=318
\mathchardef\leq=532
\mathchardef\grq=533

\catcode`\@=11 
\newcounter{pict@width}
\newcounter{pict@height}
\newlength{\pict@scale}
\newcommand{\psfigadd}[4]{%
\setcounter{pict@width}{1*\ratio{#2+\pict@scale/2}{\pict@scale}}
\setcounter{pict@height}{1*\ratio{#3+\pict@scale/2}{\pict@scale}}
\setlength{\unitlength}{\pict@scale}
\hbox to #2{\hspace{-\fill}\begin{picture}(\thepict@width,\thepict@height)
\put(0,0){\psfig{figure=#1,width=#2,height=#3,clip=}}
\SetScale{0.283466457}
\SetWidth{1.763889}
{#4}
\end{picture}}
}
\newcounter{pict@widthfst}
\newcounter{pict@widthscd}
\newcounter{pict@widthtot}
\newcommand{\psfigaddtwo}[7]{%
\setcounter{pict@widthfst}{1*\ratio{#2+\pict@scale/2}{\pict@scale}}
\setcounter{pict@widthscd}{1*\ratio{#2+#4+\pict@scale/2}{\pict@scale}}
\setcounter{pict@widthtot}{1*\ratio{#2+#4+#6+\pict@scale/2}{\pict@scale}}
\setcounter{pict@height}{1*\ratio{#3+\pict@scale/2}{\pict@scale}}
\setlength{\unitlength}{\pict@scale}
\hbox{\hspace{-\fill}\begin{picture}(\thepict@widthtot,\thepict@height)
\put(0,0){\psfig{figure=#1,width=#2,height=#3,clip=}}
\put(\thepict@widthscd,0){\psfig{figure=#5,width=#6,height=#3,clip=}}
\SetScale{0.283466457}
\SetWidth{1.763889}
{#7}
\end{picture}}
}
\newcommand{\psfigror}[4]{%
\setcounter{pict@width}{1*\ratio{#2+\pict@scale/2}{\pict@scale}}
\setcounter{pict@height}{1*\ratio{#3+\pict@scale/2}{\pict@scale}}
\setlength{\unitlength}{\pict@scale}
\hbox{\begin{picture}(\thepict@width,\thepict@height)
\put(0,\thepict@height){\psfig{figure=#1,width=#3,height=#2,clip=,angle=270}}
\SetScale{0.283466457}
\SetWidth{1.763889}
{#4}
\end{picture}}
}
\newcommand{\psfigrol}[4]{%
\setcounter{pict@width}{1*\ratio{#2+\pict@scale/2}{\pict@scale}}
\setcounter{pict@height}{1*\ratio{#3+\pict@scale/2}{\pict@scale}}
\setlength{\unitlength}{\pict@scale}
\hbox{\begin{picture}(\thepict@width,\thepict@height)
\put(0,0){\psfig{figure=#1,width=#3,height=#2,clip=,angle=90}}
\SetScale{0.283466457}
\SetWidth{1.763889}
{#4}
\end{picture}}
}
\catcode`\@=12 
\newlength\listtextwidth

\catcode`\@=11 
\newlength{\@tabfninsert}
\newlength{\@tabfnwidth}
\newcommand{\tabfootnote}[2]{%
  \setlength{\@tabfninsert}{0.8em}
  \setlength{\@tabfnwidth}{\textwidth}
  \addtolength{\@tabfnwidth}{-\@tabfninsert}
  \addtolength{\@tabfnwidth}{-0.4em}
  \noindent\makebox[\@tabfninsert][r]{\footnotesize$^{#1}$\hfil}\hfill%
  \parbox[t]{\@tabfnwidth}{\footnotesize #2\hfill}}
\catcode`\@=12 

\def\ETJ{E_{\scriptsize T}^{\rm\scriptsize jet}}
\def\ETJa{E_{\scriptsize T}^{\rm\scriptsize jet1}}
\def\ETJb{E_{\scriptsize T}^{\rm\scriptsize jet2}}
\def\ETJab{E_{\scriptsize T}^{\rm\scriptsize jet1(2)}}
\def\ETAJ{\eta^{\mbox{\rm\scriptsize jet}}}
\def\ETAJa{\eta^{\mbox{\rm\scriptsize jet1}}}
\def\ETAJb{\eta^{\mbox{\rm\scriptsize jet2}}}
\def\ETAJab{\eta^{\mbox{\rm\scriptsize jet1(2)}}}

\def\xgo{x_{\gamma}^{\mbox{\rm\tiny OBS}}}
\def\xpo{x_{p}^{\mbox{\rm\tiny OBS}}} 
\def\xpio{x_{\pi}^{\mbox{\rm\tiny OBS}}} 
\def\LN{ep\rightarrow ejjXn}
\def\inc{ep\rightarrow ejjX}
\newcommand{\myZcoosysA}{%
The ZEUS coordinate system is a right-handed Cartesian system, with the $Z$
axis pointing in the proton beam direction, referred to as the ``forward
direction'', and the $X$ axis pointing
towards the center of HERA.
The coordinate origin is at the nominal interaction point.\xspace}

\def\citeCTD{{\cite{%
nim:a279:290,*npps:b32:181,*nim:a338:254%
}}\xspace}
\def\citeCAL{{\cite{%
nim:a309:77,*nim:a309:101,*nim:a321:356,*nim:a336:23%
}}\xspace}

\begin{document}

\prepnum{DESY--09--139}

\title{
Measurement of dijet photoproduction for\\
           events with a leading neutron
           at HERA
}                                                       
                    
\author{ZEUS Collaboration}
\date{September 2009}

\abstract{
Differential cross sections for dijet photoproduction
and this process in association with a leading neutron,
$e^+ + p \rightarrow e^+ + \mathrm{jet} + \mathrm{jet} + X \, ( + n)$,
have been measured with the ZEUS detector at HERA using
an integrated luminosity of 40 pb$^{-1}$.
The fraction of dijet events with a leading neutron was studied
as a function of different jet and event variables.
Single- and double-differential cross sections are presented as a
function of the longitudinal fraction of the proton momentum carried by
the leading neutron, $x_L$, and of its transverse momentum squared, $p_T^2$.
The dijet data are compared to inclusive DIS and photoproduction
results; they are all consistent with a simple pion-exchange model.
The neutron yield as a function of $x_L$ was found to depend only on
the fraction of the proton beam energy going into the forward region,
independent of the hard process.
No firm conclusion can be drawn on the presence of rescattering
effects.
}

\makezeustitle

\def\3{\ss}

\pagenumbering{Roman}
\begin{center}                                                                                     
{                      \Large  The ZEUS Collaboration              }                               
\end{center}                                                                                       
  S.~Chekanov,                                                                                     
  M.~Derrick,                                                                                      
  S.~Magill,                                                                                       
  B.~Musgrave,                                                                                     
  D.~Nicholass$^{   1}$,                                                                           
  \mbox{J.~Repond},                                                                                
  R.~Yoshida\\                                                                                     
 {\it Argonne National Laboratory, Argonne, Illinois 60439-4815, USA}~$^{n}$                       
\par \filbreak                                                                                     
  M.C.K.~Mattingly \\                                                                              
 {\it Andrews University, Berrien Springs, Michigan 49104-0380, USA}                               
\par \filbreak                                                                                     
  P.~Antonioli,                                                                                    
  G.~Bari,                                                                                         
  L.~Bellagamba,                                                                                   
  D.~Boscherini,                                                                                   
  A.~Bruni,                                                                                        
  G.~Bruni,                                                                                        
  F.~Cindolo,                                                                                      
  M.~Corradi,                                                                                      
\mbox{G.~Iacobucci},                                                                               
  A.~Margotti,                                                                                     
  R.~Nania,                                                                                        
  A.~Polini\\                                                                                      
  {\it INFN Bologna, Bologna, Italy}~$^{e}$
% ...
\newpage                                                                                           
\enlargethispage{5cm}                                                                              
$^{\    1}$ also affiliated with University College London,                                        
United Kingdom\\                                                                                   
% ...
$^{\    5}$ also Senior Alexander von Humboldt Research Fellow at Hamburg University,              
Institute of \mbox{Experimental} Physics, Hamburg, Germany\\                                       
% ...
%
\newpage   
                                                           %
                                                           %
\begin{tabular}[h]{rp{14cm}}                                                                       
$^{a}$ &  supported by the Natural Sciences and Engineering Research Council of Canada (NSERC) \\  
$^{b}$ &  supported by the German Federal Ministry for Education and Research (BMBF), under        
          contract Nos. 05 HZ6PDA, 05 HZ6GUA, 05 HZ6VFA and 05 HZ4KHA\\                            
$^{c}$ &  supported in part by the MINERVA Gesellschaft f\"ur Forschung GmbH, the Israel Science   
          Foundation (grant No. 293/02-11.2) and the US-Israel Binational Science Foundation \\    
$^{d}$ &  supported by the Israel Science Foundation\\                                             
$^{e}$ &  supported by the Italian National Institute for Nuclear Physics (INFN) \\                
$^{f}$ &  supported by the Japanese Ministry of Education, Culture, Sports, Science and Technology 
          (MEXT) and its grants for Scientific Research\\                                          
$^{g}$ &  supported by the Korean Ministry of Education and Korea Science and Engineering          
          Foundation\\                                                                             
$^{h}$ &  supported by the Netherlands Foundation for Research on Matter (FOM)\\                   
$^{i}$ &  supported by the Polish State Committee for Scientific Research, project No.             
          DESY/256/2006 - 154/DES/2006/03\\                                                        
$^{j}$ &  partially supported by the German Federal Ministry for Education and Research (BMBF)\\   
$^{k}$ &  supported by RF Presidential grant N 1456.2008.2 for the leading                         
          scientific schools and by the Russian Ministry of Education and Science through its      
          grant for Scientific Research on High Energy Physics\\                                   
$^{l}$ &  supported by the Spanish Ministry of Education and Science through funds provided by     
          CICYT\\                                                                                  
$^{m}$ &  supported by the Science and Technology Facilities Council, UK\\                         
$^{n}$ &  supported by the US Department of Energy\\                                               
$^{o}$ &  supported by the US National Science Foundation. Any opinion,                            
findings and conclusions or recommendations expressed in this material                             
are those of the authors and do not necessarily reflect the views of the                           
National Science Foundation.\\                                                                     
$^{p}$ &  supported by the Polish Ministry of Science and Higher Education                         
as a scientific project (2009-2010)\\                                                              
$^{q}$ &  supported by FNRS and its associated funds (IISN and FRIA) and by an Inter-University    
          Attraction Poles Programme subsidised by the Belgian Federal Science Policy Office\\     
$^{r}$ &  supported by an FRGS grant from the Malaysian government\\                               
\end{tabular}                                                                                      
                                                           %
                                                           %

\newpage
                                                           %
\pagenumbering{arabic} 
\pagestyle{plain}
\section{Introduction}
\label{sec-int}

The transition of an initial-state proton  into a 
final-state neutron, $p\rightarrow n$,
has been extensively studied in hadronic
reactions~\cite{engler,robinson,flauger,
hanlon,eisenberg,blobel,abramowicz}. 
A successful phenomenological description of these results
uses Regge theory and interprets the interactions as an
exchange of virtual isovector mesons, such as
$\pi$, $\rho$, and $a_2$~\cite{sullivan,bishari,ganguli,zoller}.
At small values of the squared momentum transfer, $t$, between
the proton and the neutron, the $p\rightarrow n$ transition is expected
to be dominated by the exchange of the lightest meson, the pion.

Leading baryon processes have been previously studied in $ep$ collisions at 
HERA~\cite{
pl:b384:388,
epj:c6:587,
np:b596:3,
np:b637:3,
np:b658:3,
pl:b590:143,
epj:c41:273,
pl:b610:199,
DESY-07-011
}.
Some of these studies were performed involving a hard scale,
such as the virtuality of the photon exchanged at the lepton
vertex, $Q^2$, in deep inelastic scattering
(DIS)~\cite{epj:c6:587,pl:b384:388,np:b658:3,DESY-07-011};
the jet transverse energy, $\ETJ$,  in photoproduction of
dijets~\cite{np:b596:3};
or the charm mass in heavy-flavor production~\cite{pl:b590:143}.

Even though a hard scale is involved, the $p\rightarrow n$ transitions
are still expected to be dominated by pion exchange. 
The cross section for this type of process in $ep$ 
collisions can be written as 
\begin{equation}
\frac{d^2\sigma_{ep \rightarrow eXn}(s,x_L, t)}{dx_L\, dt} = 
f_{\pi /p}(x_L, t) \sigma_{e\pi \rightarrow eX}(s^\prime).
\label{eq-opef}
\end{equation}
This formula expresses the Regge factorization of the cross section
into the pion flux factor $f_{\pi /p}(x_L, t)$,
which describes the splitting of a proton into an $n$-$\pi$ system, 
and the cross section for electroproduction on the pion,
$\sigma_{e\pi \rightarrow eX}(s^\prime)$.
Here, $x_L$ is the fraction of the incoming
proton beam energy carried by the neutron, and $s$ and
$s^\prime = (1-x_L)s$ are the squared center-of-mass energies of
the $ep$ and of the $e\pi$ systems, respectively.

Comparisons between neutron-tagged and untagged cross sections
provide tests of the concept of vertex
factorization~\cite{pr:188:2159,*pr:d50:590}.
Under this hypothesis, the shape of the distribution of
some photon variable $V$
would neither depend on the presence of a neutron nor
explicitly on its kinematic variables $x_L$ and $t$.
Similarly, the $x_L$ and $t$ spectra
of the neutrons would be independent of the photon variable $V$.
The cross section can then be written as
\begin{equation}
\frac{d^2\sigma_{ep \rightarrow eXn}(V,x_L, t)}{dx_L\, dt} =
        g(x_L,t) G(V) ,
\label{eq-vtxfact}
\end{equation}
where $g(x_L,t)$ and $G(V)$ are  functions of the neutron and photon 
variables respectively.
The Regge factorization expressed in Eq.~\eq{opef} violates this vertex
factorization because $\sigma_{e \pi}$ has different
$s^\prime$ dependences for different processes and $s^\prime$
depends on $x_L$.
This will be further explained in \Sect{discuss}, and violations of
vertex factorization are therefore to be expected.

Rescattering effects, where the baryon interacts with the
exchanged photon~\cite{rescat3,rescat4,epj:c47:385,epj:c48:797},
are expected to increase with increasing size of the virtual photon,
i.e. decreasing $Q^2$.
This was observed in a measurement of leading neutrons in DIS and
photoproduction~\cite{DESY-07-011}.
 
\begin{figure}[htb]
\centerline{\psfig{file=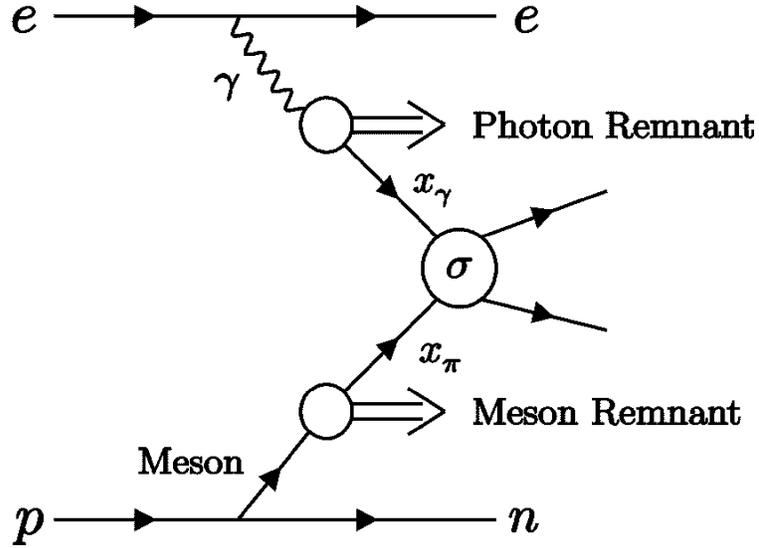,height=7.cm,width=7.cm}}
\caption{ 
Schematic of resolved photoproduction of dijets associated with a leading
neutron, mediated by meson exchange. The fraction of the 
energy of the exchanged meson (photon) 
participating in the partonic hard scattering
that produces the dijet system
is denoted by $x_{\pi}$ ($x_{\gamma}$);
the corresponding hard cross section is 
$\sigma$. In direct photoproduction, the exchanged photon
participates in the hard scattering as a point-like particle, 
there is no photon remnant, and $x_{\gamma}=1$.}
\label{fig-feynman}
\end{figure}

In high-$\ETJ$ jet photoproduction with $Q^2 \approx 0$,
two types of processes contribute to the cross section,
namely direct and resolved photon processes.
In direct  processes, the exchanged photon participates 
in the hard scattering as a point-like particle.
In resolved  processes, the photon acts as a source of partons,
one of which interacts with a parton from the incoming hadron,
see \fig{feynman}.
The more complex structure of the resolved photon may increase the probability
for the leading baryon to rescatter.
This can cause the baryon to be scattered out of the detector acceptance,
resulting in a depletion of detected baryons.
Thus, fewer leading baryons (i.e. more rescatterings) are expected in
resolved than in direct processes.

This effect was searched for, but not confirmed, in
diffractive production of dijets in photoproduction~\cite{epj:c51:549,roger}
and DIS~\cite{alessio,epj:c51:549},
where the leading proton has $x_L \approx 1$.
However, a comparison of leading neutron rates in photoproduction and DIS
showed a scale dependent suppression of
neutrons~\cite{np:b596:3,pl:b590:143,DESY-07-011};
the rates of neutrons were in good agreement
with the expectations from rescattering models~\cite{rescat3,rescat4}.

This paper reports the observation of the photoproduction of
dijets in association with a leading neutron:
\begin{equation}
e^+ + p \rightarrow e^+ + \mathrm{jet} + \mathrm{jet}+ X + n, 
\label{eq-taggedep}
\end{equation}
where $X$ denotes the remainder of the final state.
The number of events is almost an order of magnitude higher than
used for previous results~\cite{np:b596:3,epj:c41:273}.
Cross sections are presented as  functions of
the jet transverse energy, $\ETJ$,
jet pseudorapidity, $\ETAJ$,
the fraction of the photon energy carried by the dijet system, $\xgo$,
the photon-proton center-of-mass energy, $W$,
and the fraction of the proton four-momentum participating 
in the reaction, $\xpo$.
In addition, the fraction of photoproduction events with a leading neutron 
as functions of these variables is shown as a test of vertex factorization.
Finally, the $x_L$ and $p_T^2$ distributions of the leading neutrons are
shown in dijet photoproduction
and compared to similar results in DIS\cite{DESY-07-011}.

\section{Experimental setup}
\label{sec-exp}

The data sample used in this analysis was collected with the ZEUS
detector at HERA and corresponds to an integrated luminosity of 
$40\pb^{-1}$ taken during the year 2000. During this 
period, HERA operated with protons of energy $E_p=920\gev$ and positrons
of energy $E_e=27.5\gev$, yielding a center-of-mass energy of 
$\sqrt s=318\gev$.

\Zdetdesc
\Zctddesc{\footnote{\myZcoosysA\Zpsrap}}
The transverse-momentum resolution for
full-length tracks was $\sigma(p_T)/p_T=0.0058p_T\oplus0.0065\oplus0.0014/p_T$,
with $p_T$ in $\Gev$.

\Zcaldesc~
The forward-plug calorimeter (FPC)~\cite{nim:a450:235}
around the beam-pipe in the FCAL
extended calorimetry to the region $\eta \approx 4.0-5.0$.
It was a lead--scintillator calorimeter with a hadronic energy
resolution of $\sigma(E)/E = 0.65/\sqrt{E} \oplus 0.06$ ($E$ in $\gev$).

The forward neutron detectors are described in detail
elsewhere~\cite{nim:a394:121,DESY-07-011};
the main points are summarized briefly here.
The forward neutron calorimeter (FNC)
was installed in the HERA tunnel at 
\mbox{$\theta = 0^\circ$} and at $Z = 106$ m from the
interaction point in the proton-beam direction.
It was a lead--scintillator calorimeter,
segmented  vertically into towers
to allow the separation of electromagnetic and hadronic showers
by their energy sharing among towers.
The energy resolution for neutrons,
as measured in a beam test, 
was $\sigma(E_n)/E_n=0.70/\sqrt{E_n}$,
with neutron energy $E_n$ in $\gev$.
The energy scale of the FNC was determined with a
systematic uncertainty of $\pm2\%$. 
The forward neutron tracker (FNT) was installed in the FNC at a depth of
one interaction length.
It was a hodoscope designed to measure the position of neutron showers,
with two planes of scintillator fingers
used to reconstruct the $X$ and $Y$ positions of showers.
The position resolution was $\pm 0.23\cm$.
Veto counters were used to reject events in which particles had
interacted with the inactive material in front of the FNC. 
Magnet apertures limited the FNC acceptance to neutrons with 
production angles less than \mbox{$0.75$ mrad},
which corresponds to transverse momenta
$p_T < E_n\theta_{\mbox{\rm\tiny max}}=0.69 \, x_L\gev$.

The luminosity was determined from the rate of the bremsstrahlung process,
$ep \rightarrow e \gamma p$, where the photon was measured with a 
lead--scintillator calorimeter~\cite{desy-92-066,acpp:b32:2025} 
located at \hfill\\ \mbox{$Z = -107$ m}.

\section{Data selection and kinematic variables}
\label{sec:ev-sel}

A three-level trigger system was used to select events
online~\cite{zeus:1993:bluebook,uproc:chep:1992:222}.
At the second level, cuts were made to reject beam-gas
interactions and cosmic rays.
At the third
level, jets were reconstructed using the energies and positions of the
CAL cells. Events with at least two jets with transverse
energy in excess of $4.5$~GeV and $|\ETAJ|$ below $2.5$ were
accepted. No requirement on the FNC was made at any trigger level.

Offline, tracking and calorimeter information were combined to form energy-flow
objects (EFOs)~\cite{epj:c1:81,thesis:briskin:1998}. The $\gamma p$ 
center-of-mass energy, $W$, was reconstructed using the Jacquet-Blondel
method~\cite{proc:epfacility:1979:391} as
\mbox{$W_{\rm {JB}}=\sqrt{y_{\rm {JB}}s}$}, where
\mbox{$y_{\rm {JB}} = \sum_i (E_i - E_{Z,i}) /2E_e$} is an estimator
of the inelasticity variable $y$, and
\mbox{$E_{Z,i} = E_i\cos\theta_i$}; $E_i$ is the energy of EFO
$i$ with polar angle $\theta_i$.
The sum runs over all EFOs. The energy $W_{\rm {JB}}$ was corrected
for energy losses using the Monte Carlo (MC) samples described in 
Section~4. After corrections, the sample was restricted to
\mbox{$130 < W < 280$ GeV}. Events with a reconstructed positron candidate
in the main detector were rejected. The selected photoproduction sample
consisted of events from $ep$ interactions with $Q^2<1$~GeV$^2$ and a
mean $Q^2\approx 10^{-3}$~GeV$^2$.

The $k_T$ cluster algorithm~\cite{np:b406:187} was used in the
longitudinally invariant inclusive mode~\cite{pr:d48:3160} to
reconstruct jets in the measured hadronic final state from
the energy deposits in the CAL cells (calorimetric jets). The axis of
the jet was defined according to the Snowmass
convention~\cite{proc:snowmass:1990:134}. The jet search was performed
in the $(\eta-\phi)$ plane of the laboratory
frame. Corrections~\cite{pl:b560:7} to the jet transverse energy,
$\ETJ$, were applied as a function
of the jet pseudorapidity, $\ETAJ$, and $\ETJ$,
and averaged over the jet azimuthal angle.
Events with at least two jets of
$\ETJab>7.5(6.5)\gev$, 
where $\ETJab$ is the transverse energy
of the highest (second highest) $\ETJ$ jet,
and $-1.5<\ETAJ<2.5$, were retained.

Leading neutron events were selected from the dijet sample by applying
criteria described previously~\cite{DESY-07-011}.
The main requirements are listed here.
Events were required to have energy deposits
in the FNC with energy $E_{\rm FNC} > 184 \gev$ ($x_L>0.2)$
and timing consistent with the triggered event.
In addition the deposits had to be close to the zero-degree point
in order to
reject protons bent into the FNC top section. Electromagnetic showers
from photons were rejected by requiring the energy sharing among the
towers to be consistent with a hadronic shower. Showers which started
in dead material upstream of the FNC were rejected by requiring that
the veto counter had a signal of less than one mip.
Additional information from the FNT was used to select a subsample of events
where a good position and thus $p_T^2$ measurement was possible.
The channel with the largest pulse-height in each of the hodoscope planes was
required to be above a threshold to select neutrons which
showered in front of the FNT plane, and transverse shower profiles
were required to have only one peak to minimize
the influence of shower fluctuations.

After the requirements described above,
the final dijet sample contained 583168 events,
of which a subsample of 9193 events had a neutron tag,
and 4623 of these also had a well measured neutron position.

The fractions of the photon and proton four-momenta entering the hard
scattering, $x_{\gamma}$ and $x_p$ respectively,
were reconstructed via
\begin{equation}
\xgo = \frac{\ETJa e^{-\ETAJa}
                       + \ETJb e^{-\ETAJb}}
                                              {2 E_e y_{\rm JB}},
\label{eq-xgo}
\end{equation}
\begin{equation}
\xpo = \frac{\ETJa e^{\ETAJa}
                  + \ETJb e^{\ETAJb}}{2E_p},
\label{eq-xpo}
\end{equation}
where $\ETAJab$ and $\ETJab$ are the pseudorapidity 
and transverse energy, respectively, of the highest
(second highest) $\ETJ$ jet.
The observable $\xgo$ was used to separate the
underlying photon processes since it is small (large) for
resolved (direct) processes.
The fraction of the exchanged pion four-momentum entering the hard
scattering, $x_{\pi}$ in \fig{feynman}, was reconstructed as
$ \xpio =  \xpo / (1 - x_L)$.

\section{Monte Carlo simulations}
\label{sec-montecarlo}

\subsection{Detector corrections}
\label{sec-mcdetcor}

Samples of MC events were generated to study the
response of the central detector to jets of hadrons 
and the response of the forward neutron detectors.
The acceptances of the central and forward detectors are independent
and the overall acceptance factorizes as the product of the two;
they were evaluated using two separate MC programs.

The programs {\sc Pythia}~6.221~\cite{pythia} and
{\sc Herwig}~6.1~\cite{cpc:67:465,*jhep:0101:010} were used to
generate photoproduction events for resolved and direct processes
producing dijets in the central detector.
Fragmentation into hadrons was performed
using the Lund string model~\cite{prep:97:31} as implemented in 
{\sc Jetset}~\cite{cpc:82:74,*cpc:135:238,cpc:39:347,*cpc:43:367} in
the case of {\sc Pythia}, and a cluster model~\cite{np:b238:492} in
the case of {\sc Herwig}. 
The generated events were passed through the
{\sc Geant}~3.13-based~\cite{tech:cern-dd-ee-84-1} ZEUS detector- and
trigger-simulation programs~\cite{zeus:1993:bluebook}. They were
reconstructed and analyzed by the same program chain as the data.

The {\sc Pythia} program was used to determine the central-detector
acceptance corrections.
Samples of resolved and direct processes were generated separately.
The resolved sample was reweighted as a function of  $x_\gamma$
and the direct sample as a function of $W$.
The reweighting and relative contributions of
the two samples were adjusted to give the best description
of the measured $x_\gamma$ and $W$ distributions.
Different reweighting and mixing factors were applied
for the inclusive and neutron-tagged jet samples.

The {\sc Herwig} program was used to check the systematic
effects of the detector corrections.
Direct and resolved photon processes were generated with
default parameters and multiple interactions turned on.

A detailed description of the efficiencies and correction factors for
the leading neutron measurements is given elsewhere~\cite{DESY-07-011}.

\subsection{Model comparisons}
\label{sec-mcmodels}

Previous studies have shown that MC models generating leading
neutrons from the fragmentation of the proton remnant do not
describe the neutron $x_L$ and $p_T^2$ distributions in
DIS nor in photoproduction~\cite{DESY-07-011}.
Models incorporating pion exchange gave the best description of
the leading neutrons;
also models with soft color interactions (SCI)~\cite{pl:b366:371}
were superior to the fragmentation models.
Monte Carlo programs incorporating these non-perturbative processes
were used for comparison to the present dijet photoproduction data.

The {\sc Rapgap} model incorporates pion exchange to simulate
leading baryon production.
It also includes Pomeron exchange to simulate diffractive events.
These processes are mixed with standard fragmentation
according to their respective cross sections.
The PDF parameterizations used were
CTEQ5L~\cite{epj:c12:375} for the proton,
the GRV-G LO~\cite{pr:d45:3986} for the photon,
the H1 fit 5~\cite{zfp:c76:613} for the Pomeron
and GRV-P LO fit~\cite{zfp:c53:651} for the pion.
The light-cone exponential flux factor~\cite{pl:b338:363}
was used to model pion exchange.

The SCI model assumes that soft color exchanges give variations in the
topology of the confining color-string fields which then
hadronize into a final state which can include a leading neutron. 
It was interfaced to the {\sc Pythia} program~\cite{epj:c33:542};
this implementation of {\sc Pythia} did not include
multiple parton interactions.

\section{Systematic uncertainties}
\label{sec-system}

Systematic uncertainties associated with the CTD and the CAL influence the
jet measurement;
those associated with the FNC influence the neutron measurement.
They are considered separately.

For the jet measurements, 
the systematic effects are grouped into the following 
classes, their contributions to the uncertainties on
the cross sections being given in parentheses:
\begin{itemize}
  \item knowledge of absolute jet energy scale to 3\%: (1--6\%);
  \item model dependence:
        the acceptances were estimated
        using {\sc Herwig} instead of
        {\sc Pythia} tuned as described in the previous section
        (5--9\%);
  \item event selection: 
        variation of  $W$ and $\ETJ$ cuts by one standard deviation
        of the resolution
        (1--6\% each for $W$ and $\ETJ$).
\end{itemize}
Together, these effects resulted in uncertainties of 7--15\% on the
jet cross sections.
The overall normalization has an additional uncertainty of 2.25\%
due to the uncertainty in the luminosity  measurement.

An extensive discussion of the systematic effects related to  
the neutron measurement is given elsewhere \cite{DESY-07-011};
the effects are summarized here.
The neutron acceptance is affected by uncertainties in the
beam zero-degree point and the dead material map,
and uncertainties in the $p_T^2$ distributions which enter into
the computation of the neutron acceptance.
The 2\% uncertainty on the FNC energy scale also
affects the $x_L$ and $p_T^2$ distributions.
Systematic uncertainties from these effects were typically
5--10\% of the measured quantities,
for example the exponential $p_T^2$ slopes.
The systematic variations largely affect the neutron acceptance
and result in a correlated shift of neutron yields.
Corrections for efficiency of
the cuts and backgrounds in the leading neutron sample were applied to the
normalization of the neutron yields.
The corrections accounted for veto counter over- and
under-efficiency and neutrons from proton beam-gas interactions.
The overall systematic uncertainty on the normalization of the
neutron cross sections from these corrections was $ \pm 2.1 \%$.
Combined with the other neutron systematics, the overall
systematic uncertainty on the total neutron rate was  $ \pm 3 \%$.

\section{Results}
\label{sec-results}

\subsection{Jet cross sections and ratios}
\label{sec-xsec}

The inclusive dijet and neutron-tagged dijet photoproduction cross sections
have been measured for jets with
\mbox{$\ETJab>7.5(6.5)\gev$} and
\mbox{$-1.5<\ETAJ<2.5$},
in the kinematic region 
\mbox{$Q^2<1\gev^2$} and \mbox{$130<W<280\gev$},
with the additional restriction of
\mbox{$x_L>0.2$} and \mbox{$\theta_n<0.75$ mrad}
for the neutron-tagged sample.
The fraction of dijet events with a leading neutron,
the yield $r_{\rm LN}$, in the measured kinematic region is
\begin{equation}
r_{\rm LN} = \frac{\sigma_{\LN}}{\sigma_{\inc}} =
  6.63 \pm 0.07 \, {\rm (stat.)} \, \pm{0.20} \, {\rm
  (syst.)} \%.
\label{eq-rln}
\end{equation}
In this ratio, most of the systematic effects of the
dijet selection cancel, and the uncertainty is dominated
by the systematic effects of the neutron selection.

\begin{figure}[htb]
\centerline{\psfig{file=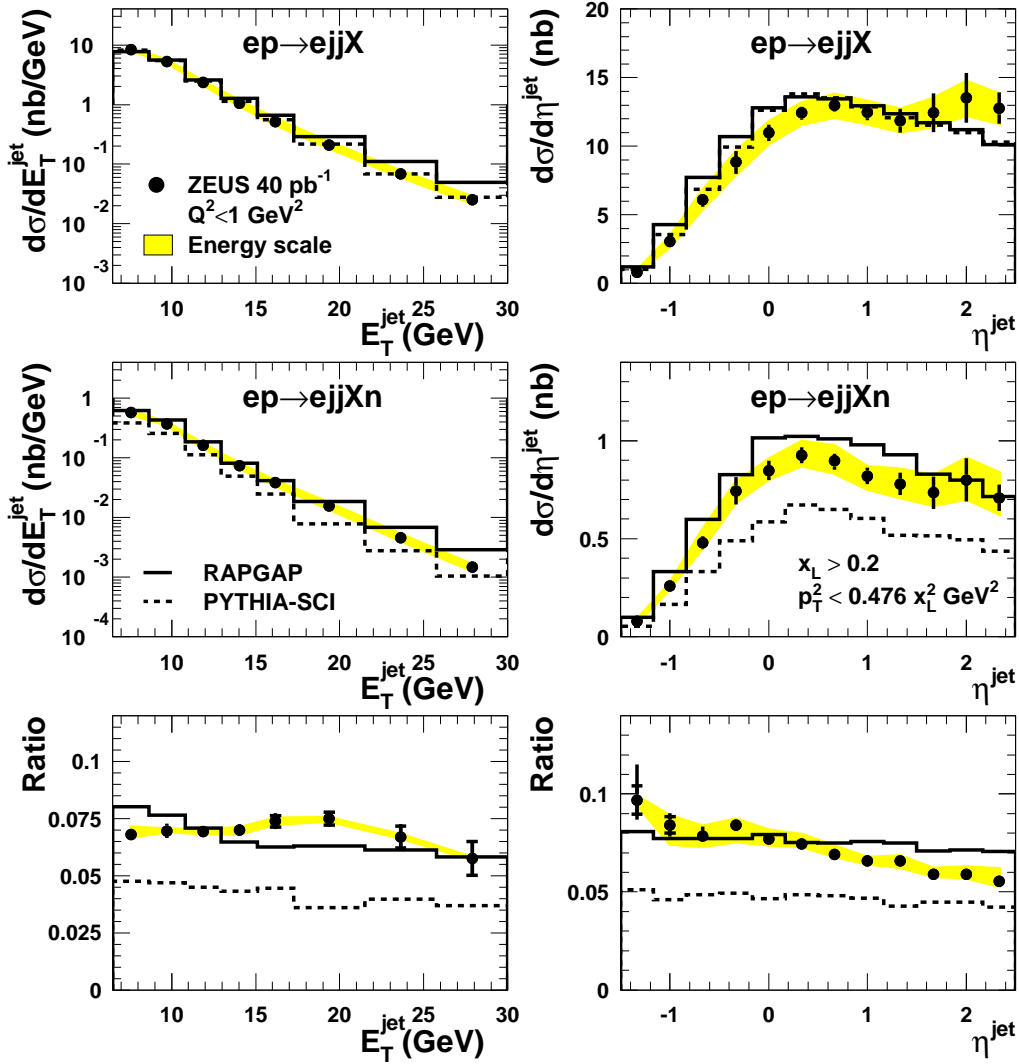,height=15.cm} }
\caption{
Differential neutron-tagged and untagged dijet photoproduction
cross sections as functions of $\ETJ $ and $\eta$.
The ratios between cross sections, the neutron yields, are also given. 
The inner error bars, where visible, show the statistical uncertainty;
the outer error bars, where visible, show the statistical and jet-related 
systematic uncertainties other than CAL energy scale summed in quadrature;
the shaded bands show the contribution to the latter from the CAL energy scale.
There is an overall systematic uncertainty on the normalization of the
neutron cross-sections and the ratios
of  $ \pm 3 \%$ which is not shown. 
An overall uncertainty on the normalization of the cross sections 
of 2.25\%  due to the luminosity  measurement is also not shown. 
The histograms show the predictions of the Monte Carlo models
{\sc Rapgap} (solid histogram) and {\sc Pythia} with SCI (dashed histogram)
as described in the text. 
}
\label{fig-eteta}
\end{figure}

The differential cross sections  for neutron-tagged and untagged events
as functions of the jet variables $\ETJ$ and $\ETAJ$ are presented in
\fig{eteta} and summarized in \tab{eteta}.
They contain two entries per event, one for each jet.
Also shown are the neutron yields $r_{\rm LN}$ as defined
in Eq.~\eq{rln} as a function of the relevant variable.
The cross sections as functions of $\ETJ$ show a reduction of about
three orders of magnitude within the measured range.
The neutron yield is approximately constant as a
function of $\ETJ$.
The cross sections as functions of $\ETAJ$ rise over
the range $-1.5<\ETAJ<0.5$; for higher values of $\ETAJ$ they flatten.
The neutron yield decreases with $\ETAJ$.

\Fig{eteta} also shows the predictions of the {\sc Rapgap}
and SCI programs implemented as described in \Sect{mcmodels}.
Both are close in magnitude to the inclusive data.
They both describe the steep drop with $\ETJ$ 
and the shape of the $\ETAJ$ distributions.
For neutron-tagged events
{\sc Rapgap} slightly overestimates and the SCI model
clearly underestimates the cross section.
They underestimate the decrease of the neutron yield with $\ETAJ$.

\begin{figure}[htb]
\centerline{\psfig{file=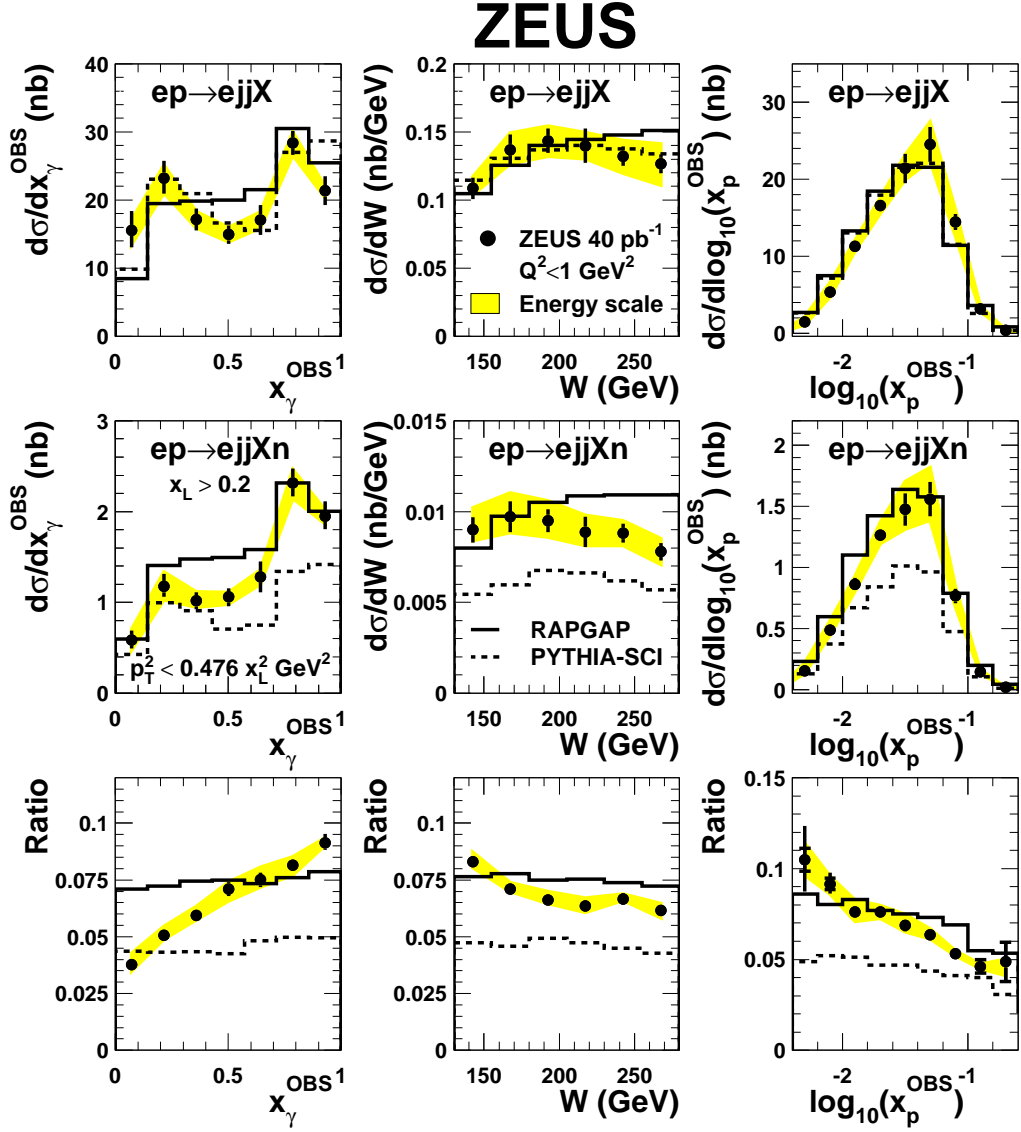,height=15.cm}}
\caption{
Differential neutron-tagged and untagged dijet photoproduction
cross sections as functions of $\xgo$, $W$ and $\log_{\rm 10}(\xpo)$.
The ratios between between cross sections, the neutron yields, are also given. 
Details are as in Fig. \ref{fig-eteta}.
}
\label{fig-xgamwxp}
\end{figure}

The differential cross sections as functions of 
the event variables $\xgo$, $W$ and $\xpo$ are 
presented in \fig{xgamwxp} and summarized in \tab{xgamwxp}.
The cross sections as functions of $\xgo$ show two peaks at 
$\xgo\approx 0.2$ and $\xgo\approx 0.8$ which can be attributed to 
the resolved- and direct-photon contributions, respectively.
The neutron-tagged sample has a significantly smaller
resolved contribution at low $\xgo$. This is seen clearly
in the yield which rises by a factor of two from low to high $\xgo$.
The cross sections are roughly flat as a function of $W$;
the yield exhibits a mild decrease with increasing $W$.
The measured range of $\xpo$ is 0.04 to 0.25 and
the cross section peaks close to $\xpo = 0.05$.
The neutron yield decreases by a factor of two across the range measured.

Also shown in \fig{xgamwxp} are the predictions of the {\sc Rapgap}
and SCI models.
{\sc Rapgap} does not have a two-peaked structure
as a function of $\xgo$, whereas the SCI model predicts the
drop in cross section at central values of $\xgo$ exhibited by the data.
For the neutron-tagged sample, {\sc Rapgap} overestimates the
cross section in the resolved regime while SCI underestimates
the cross section in the direct regime.
Both models predict the relatively weak dependence of the
cross section on $W$
and describe reasonably well the shape of the $\xpo$ distribution.
Neither model can reproduce the dependence of the neutron yield
on $\xgo$ and $W$.
The {\sc Rapgap} model predicts a small decrease of the neutron
yield with $\xpo$. However, the decrease is more pronounced in the data.
The SCI model does not reproduce this feature at all.

\begin{figure}[t]
\centering
\epsfig{file=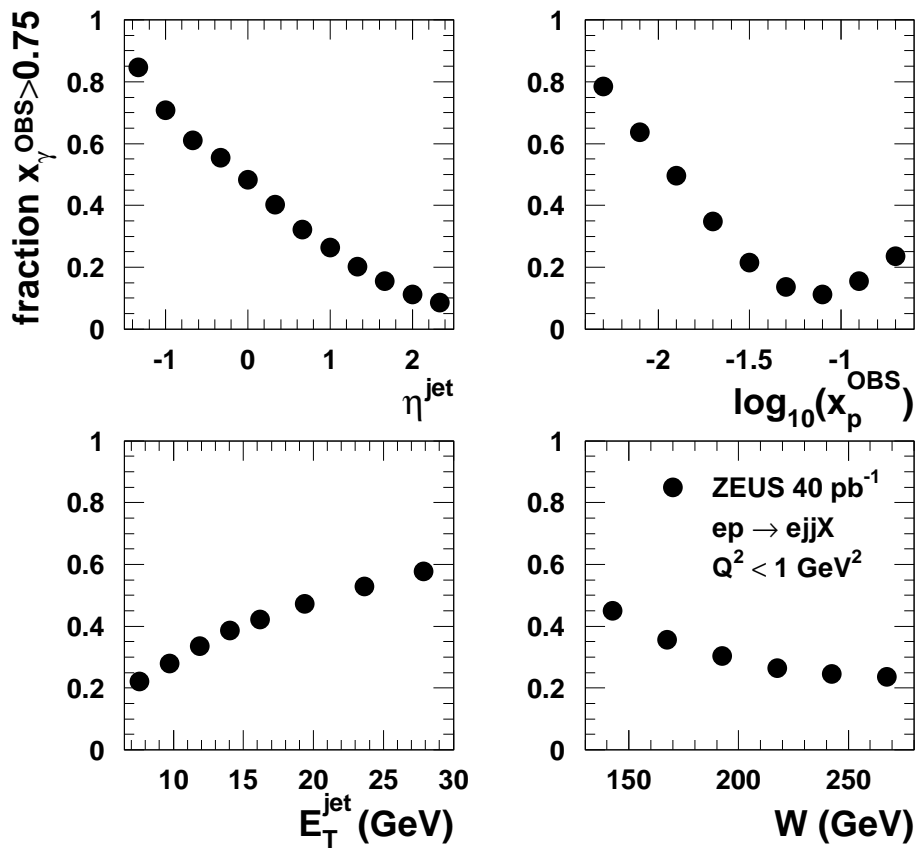,width=14cm}
\caption{
Direct photon contributions ($\xgo > 0.75$)
as functions of the other jet and event variables.
Statistical uncertainties are smaller than the plotted solid points.
}
\label{fig-fdirect}
\end{figure}

The dependence of the neutron yield on $\ETAJ$, $\xgo$ and $\xpo$ as seen
in \figand{eteta}{xgamwxp} indicates a violation of vertex factorization.
This might be explained by the Regge factorization as discussed in \Sect{int}.
The factorization violations seen in different variables are connected.
A strong anticorrelation between the direct contribution $(\xgo>0.75)$
and $\ETAJ$ and $\xpo$ is apparent in the data in \fig{fdirect}.
Events with low values of these variables contain up to 80\% direct
component, events with high values contain up to 90\% resolved component.
The observed drop of neutron yields at high $\ETAJ$ and $\xpo$
can thus be accounted for by a lower neutron yield in the
resolved photon contribution.
The smaller dependence of the neutron yield on $\ETJ$ and $W$
is consistent with this mechanism.

The H1 collaboration has also reported similar
measurements~\cite{epj:c41:273}.
They were made in a similar region of $\ETJ$, $\ETAJ$
and $W$ as the present analysis, but restricted to $x_L>0.61$.
The same pattern of vertex factorization violation was observed there.
Also, after accounting for the different $x_L$ ranges,
the cross sections are consistent.

\subsection{Neutron $\mathbf{x_L}$ distribution and pion structure}
\label{sec-xl}

\begin{figure}[htb]
\centerline{\psfig{file=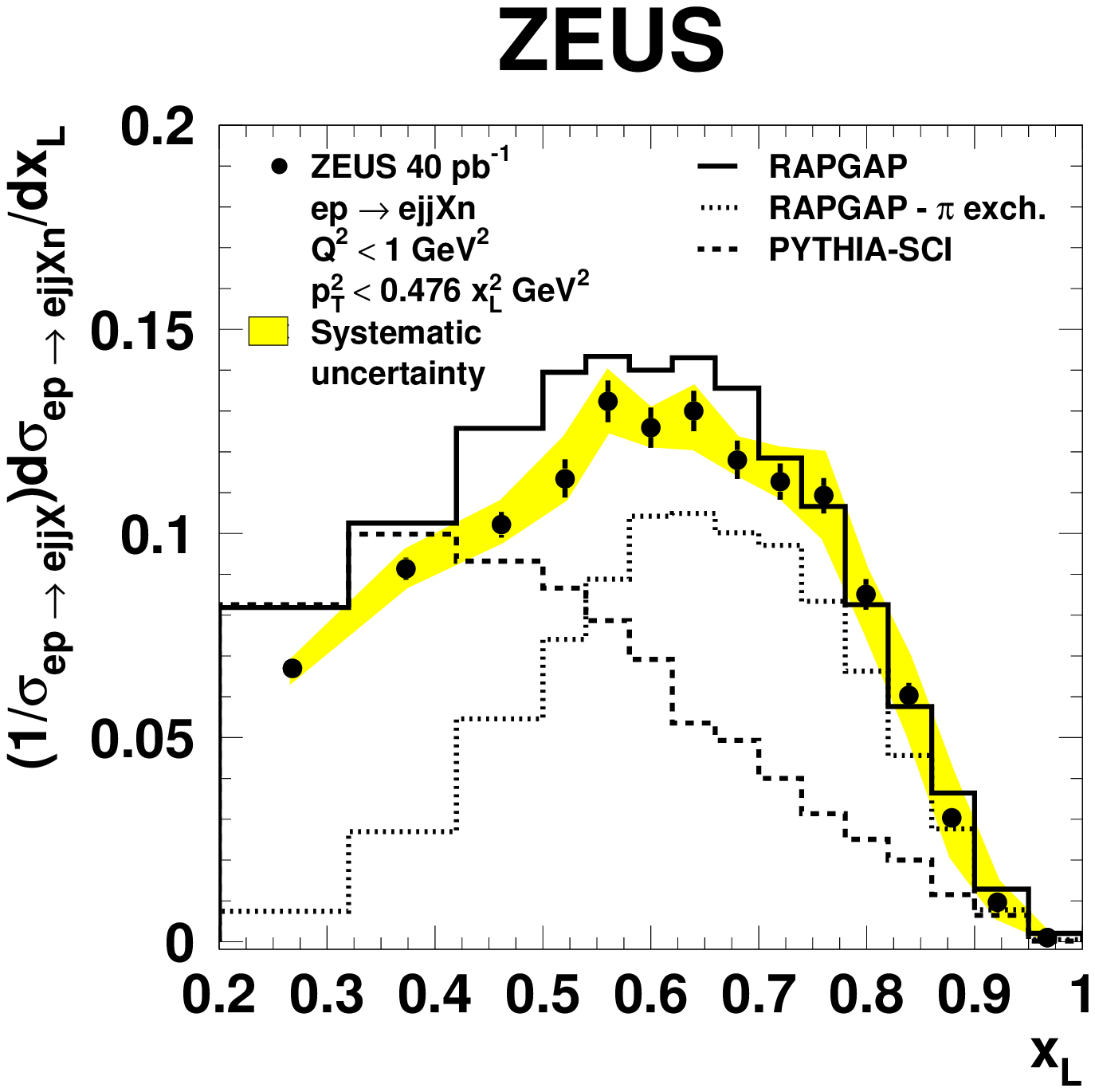,width=14cm}}
\caption{
The normalized differential distribution 
\mbox{$(1/\sigma_{\inc})d\sigma_{\LN}/dx_L$}
in dijet events.  
The error bars show the statistical uncertainty;
neutron-related systematic uncertainties are shown separately as a 
shaded band. An overall 
systematic uncertainty on the normalization of the
neutron cross-sections 
of  $ \pm 2.1 \%$ is not shown. 
The solid histogram shows the prediction of the
full {\sc Rapgap} model; the dotted histogram is the
contribution from pion exchange.
The dashed histogram is the prediction of {\sc Pythia} with SCI.
}
\label{fig-xlthcut}
\end{figure}

\Fig{xlthcut} 
shows the normalized differential cross-section
$ (1/\sigma_{\inc}) d\sigma_{\LN}/dx_L $
for neutrons with  $\theta_n < 0.75 \mrad$, which
corresponds to $p_T^2 < 0.476 \, x_L^2 \gev^2$.
The distribution rises from the lowest $x_L$
values due to the increase in $p_T^2$ phase space.
It reaches a maximum for $x_L\approx 0.6$,
and falls to zero at the endpoint $x_L=1$.
Also shown are the predictions of the MC models.
The {\sc Rapgap} program gives a fair description of both the shape
and normalization of the data, although its prediction is significantly
above the data for $x_L<0.7$.
The SCI model does not describe the data, predicting too
few events with neutrons and with a spectrum peaked at too low $x_L$.
Also shown in \fig{xlthcut} is the pion-exchange contribution
to the {\sc Rapgap} prediction for the $x_L$ distribution.
This contribution is essential for the {\sc Rapgap}
prediction to describe the measured distribution.
It dominates for $x_L>0.6$. Thus, in this region the
dijet photoproduction data are sensitive to the pion structure.

\begin{figure}[htb]
\centerline{\psfig{file=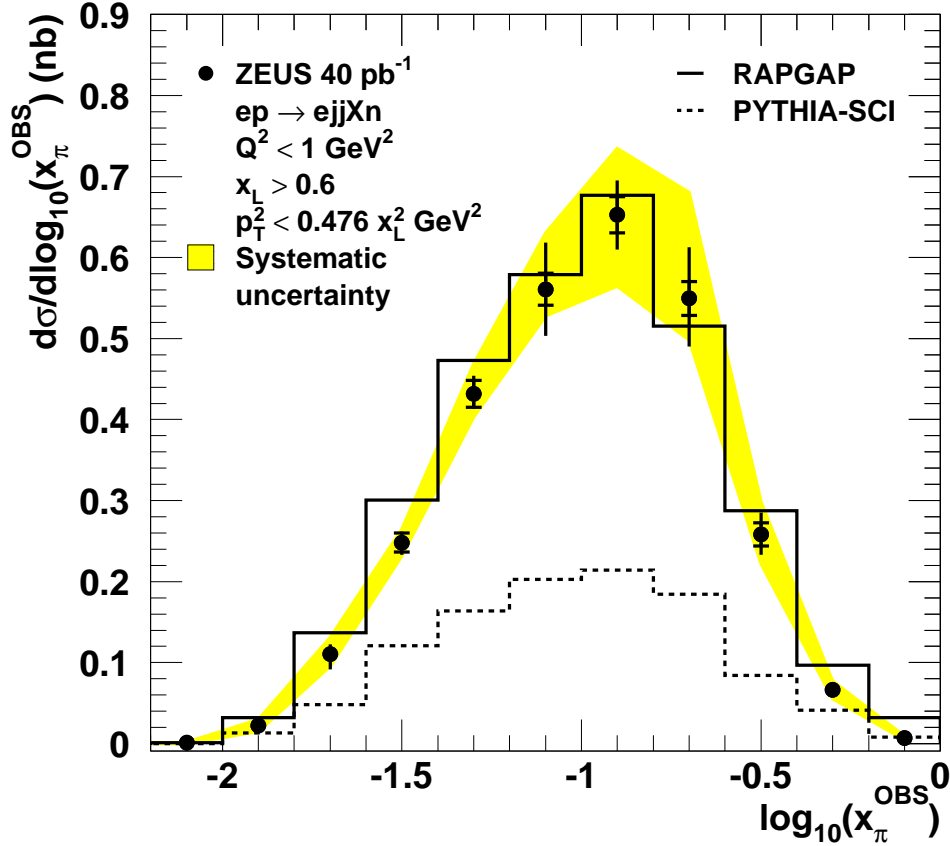,width=14cm}}
\caption{
Differential cross section for $x_L>0.6$ as a function
of $\log_{\rm 10}(\xpio)$,
the fraction of the exchanged pion's momentum participating in the
production of the dijet system for the neutron-tagged sample. 
Details are as in Fig. \ref{fig-eteta}.
The $x_L$ cut restricts the sample to the 
region where pion exchange is the dominating process.
}
\label{fig-xpion}
\end{figure}

\Fig{xpion} shows the neutron cross section as a function of
$\log_{10}(\xpio)$
for $x_L>0.6$; the values are listed in \tab{xpion}.
The range in $\xpio$ is from $0.01$ to $0.6$;
the distribution peaks near $\xpio \approx 0.13$.
Also shown in \fig{xpion} are the predictions
of {\sc Rapgap} and SCI.
The former provides a good description of the data while
the latter underestimates the cross section by about a factor of three.
It should be noted that {\sc Rapgap}, using the pion PDF parameterization
GRV-P LO~\cite{zfp:c53:651} based on fixed-target data
with $x_{\pi}>0.1$, is able to describe the cross section
down to $x_{\pi} \approx 0.01$.

\subsection{$\mathbf{p_T^2}$ distributions}
\label{sec-ptsq}

\begin{figure}[htb]
\centerline{\psfig{file=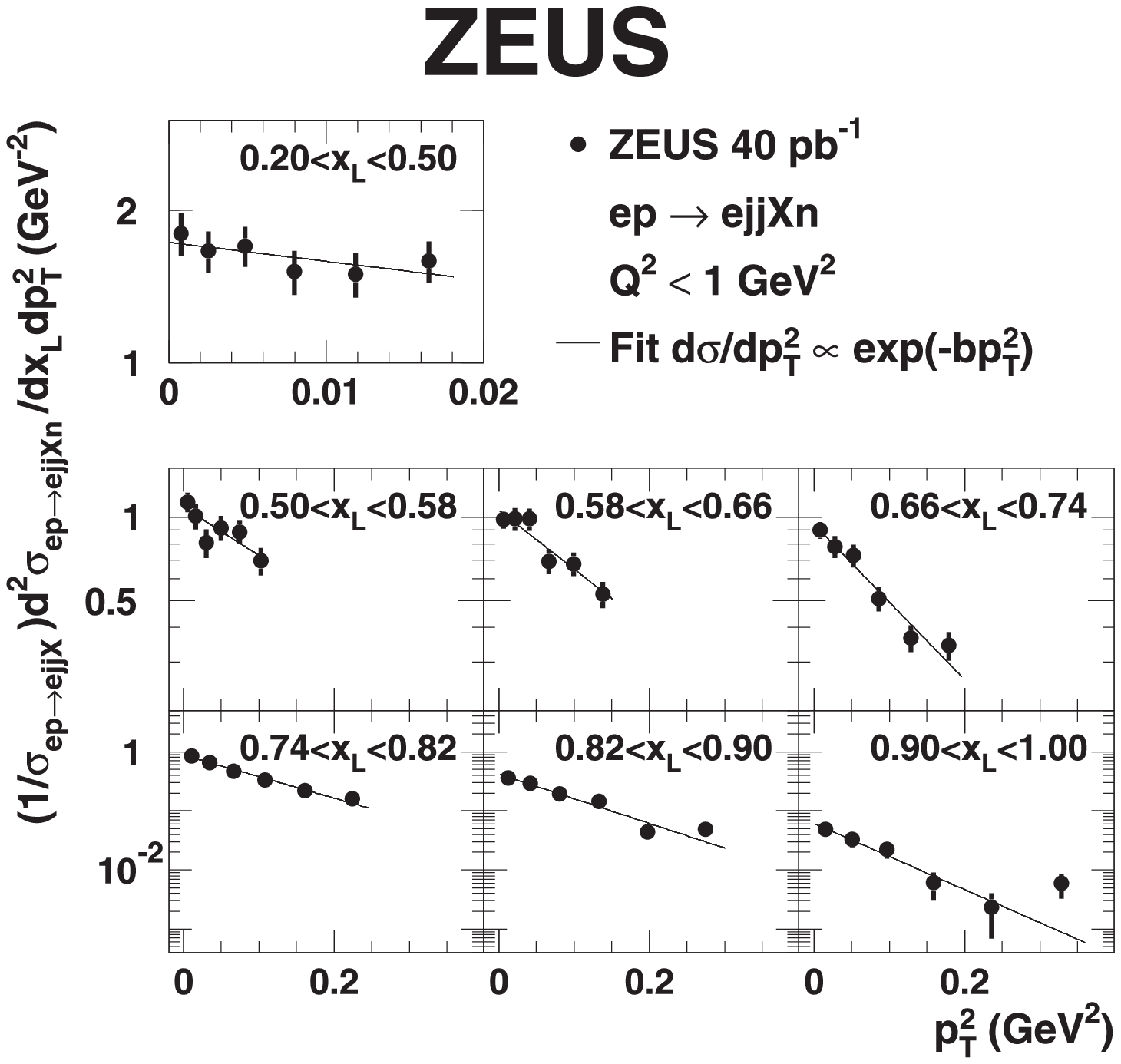,height=14.cm}}
\caption{
The $p_T^2$ distributions in bins of $x_L$.
The statistical uncertainties are shown by vertical error bars;
in some cases they are smaller than the plotted symbol. 
The systematic uncertainties are not shown.
The line on each plot is the result of a fit to the
form \mbox{$d\sigma_{\LN}/dp_T^2 \propto \exp(-bp_T^2)$}.
}
\label{fig-ptsq}
\end{figure}

\begin{figure}[htb]
\centerline{\psfig{file=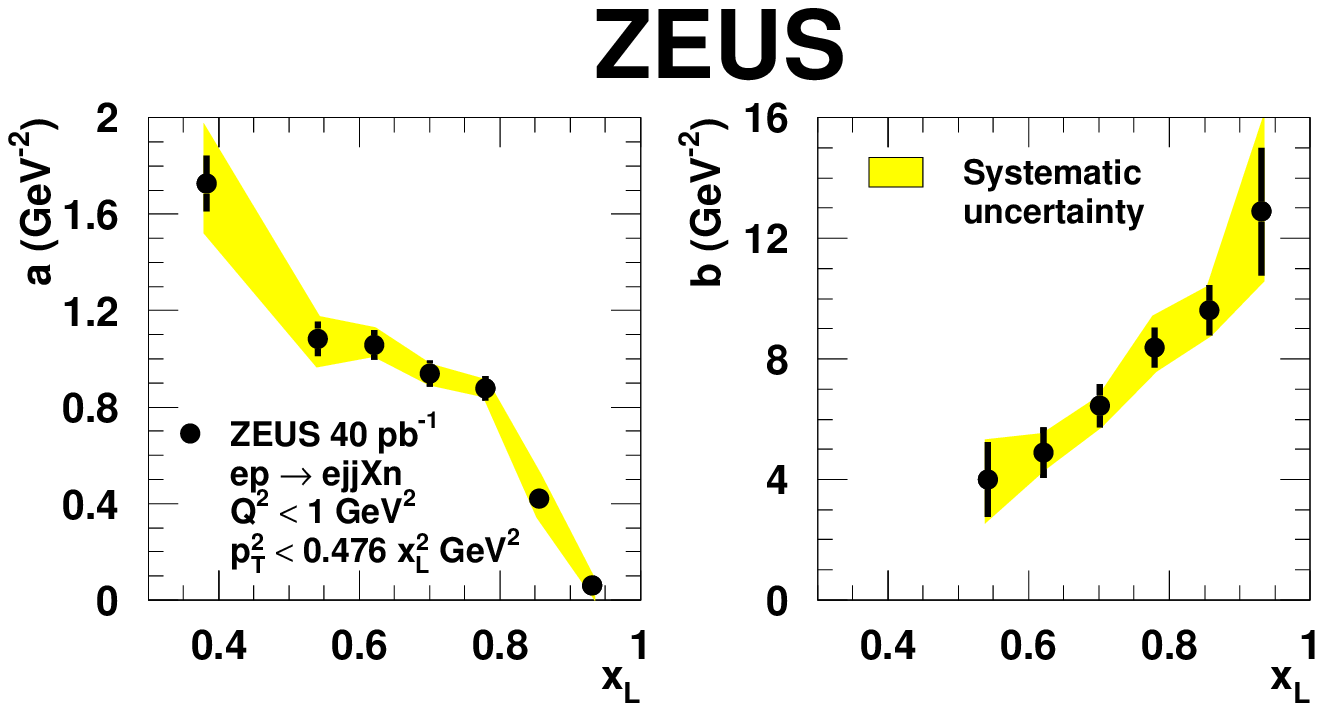,width=15.5cm}}
\caption{
Intercepts $a$ and exponential slopes $b$ versus $x_L$
from fits of the $p_T^2$ distributions to the form
\mbox{$(1/\sigma_{\inc})d^2\sigma_{\LN}/dx_L dp_T^2 = a \exp(-bp_T^2)$}
over the range $p_T^2 < 0.476 x_L^2 \gev^2$.
The error bars show the statistical uncertainties;
the shaded bands show the neutron-related systematic uncertainties.
The band for the intercepts does not include the overall
normalization uncertainty of $\pm 2.1\%$.
}
\label{fig-intslope}
\end{figure}

The  $p_T^2$  distributions of the leading neutrons in different $x_L$ bins
are shown in \fig{ptsq} and summarized in \tab{ptsq}.
They are presented as normalized doubly differential distributions,
$ (1/\sigma_{\inc}) d^2 \sigma_{\LN}/dx_L dp_T^2$.
The bins in $p_T^2$ are at least as large as the
resolution, which is dominated by the $p_T$ spread of the proton beam.
The varying $p_T^2$ ranges of the data are due
to the aperture limitation.
The line on each plot is a fit to the functional form
$d\sigma_{\LN}/dp_T^2 \propto \exp(-bp_T^2)$.
Each distribution is compatible with a single
exponential within the statistical uncertainties.
Thus, with the parameterization
\begin{equation}
 \frac{1}{\sigma_{\inc}} \frac{d^2 \sigma_{\LN}}{dx_L dp_T^2} =
               a(x_L) \:  e^{\displaystyle -b(x_L) p_T^2} ,
\label{eq-doublediff}
\end{equation}
the neutron $(x_L,p_T^2)$ distribution is
characterized by the slopes $b(x_L)$ and intercepts
$ a(x_L) =
\left. (1/\sigma_{\inc}) d^2 \sigma_{\LN}/dx_L
dp_T^2\right|_{p_T^2=0} $.
The results of exponential fits in bins of $x_L$ 
for the intercepts and the slopes are shown in
\fig{intslope} and summarized in \tab{intslope}.
The systematic uncertainties were evaluated by
making the variations discussed in \Sect{system}
and repeating the fits.
The intercepts fall rapidly from the lowest $x_L$,
drop mildly in the region $x_L=0.5-0.8$,
and fall to zero at the endpoint $x_L=1$.
In the lowest $x_L$ bin, the slope is consistent with zero
and is not plotted;
above $x_L=0.5$ the slope rises roughly linearly to a value of
$b \approx 13 \gev^{-2}$ at $x_L=0.93$.

\subsection{Comparisons of different processes}
\label{sec-compare}

\subsubsection{Comparison to neutron production in DIS}
\label{sec-disdij}

\begin{figure}[t]
\centering
\epsfig{file=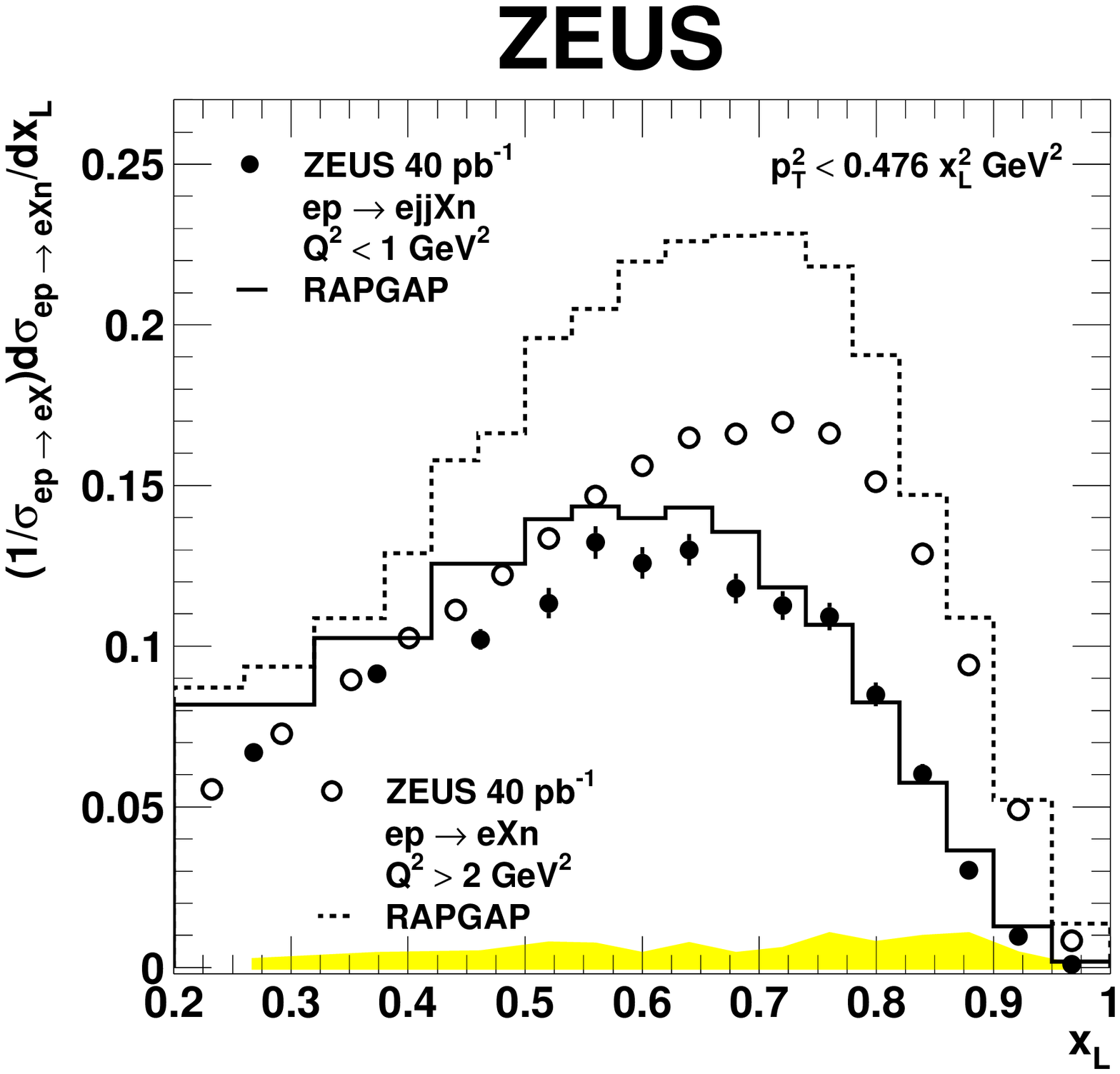,width=14cm}
\caption{
Normalized differential distributions
\mbox{$(1/\sigma_{ep\rightarrow eX})d\sigma_{ep\rightarrow eXn}/dx_L$}.
The solid points are for dijet photoproduction and
the open points for DIS{\protect \cite{DESY-07-011}}.
Both distributions are normalized by
their respective untagged cross sections.
Statistical uncertainties are shown as vertical bars; in the DIS case they
are smaller than the plotted symbols. The systematic uncertainties,
shown as the shaded band, are similar for both data sets.
The histograms are the predictions of {\sc Rapgap}.
}
\label{fig-xldijdis}
\end{figure}

\Fig{xldijdis} shows the normalized $x_L$ distribution
of leading neutrons in dijet photoproduction and in inclusive DIS
with $Q^2>2\gev^2$~\cite{DESY-07-011}.
The yield of neutrons from dijet photoproduction agrees with that in DIS
at low $x_L<0.4$, but is lower at higher $x_L$.
For $x_L>0.8$ the yield in dijet photoproduction is more than
a factor of two lower than in inclusive DIS.

\Fig{xldijdis} also shows the predictions of {\sc Rapgap}
for dijet photoproduction and DIS.
The predicted shapes are in fair agreement with the measurements.
However, the predicted neutron yield is $\approx 10\%$ too high for
dijet photoproduction and $\approx 30\%$ too high for DIS.
The shapes of the distributions for the two processes are
compared using the ratio
\begin{equation}
\rho = \frac{
(1/\sigma_{\inc}) d\sigma_{\LN}/dx_L}
{(1/\sigma_{ep\rightarrow eX})d\sigma_{ep\rightarrow eXn}/dx_L
 (Q^2 > 2 \gev^2)} \, .
\label{eq-rho}
\end{equation}
The result is shown in \fig{ratiojjdis}.
After normalizing each prediction to its respective data set,
{\sc Rapgap} provides a fair description of the
drop of the neutron yield with $x_L$ in dijet photoproduction
relative to that in DIS.

\begin{figure}[t]
\centering
\epsfig{file=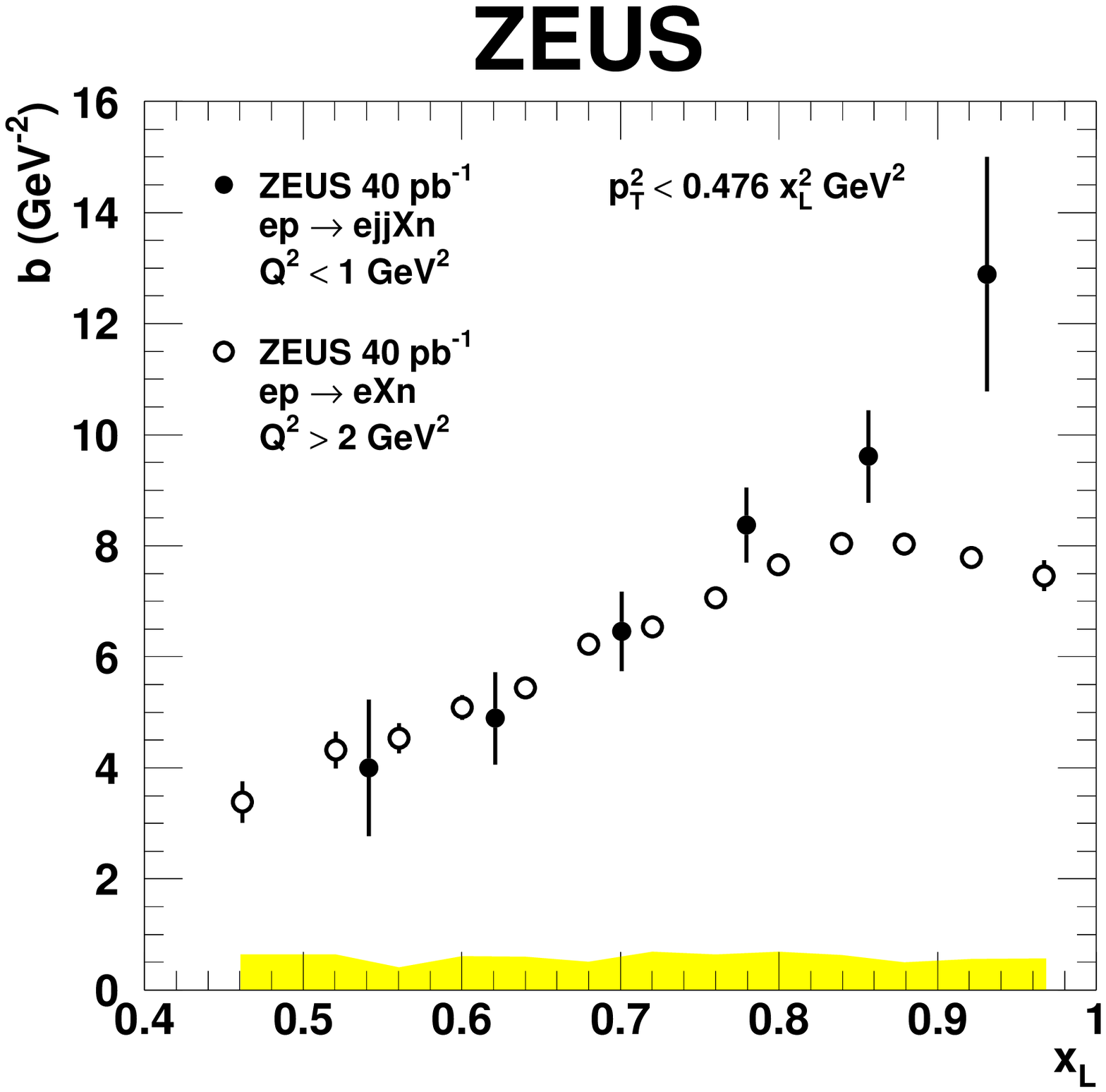,width=14cm}
\caption{
Exponential slopes $b$ versus $x_L$ from fits of the $p_T^2$
distributions to the form
$d\sigma_{ep\rightarrow eXn}/dp_T^2 \propto \exp(-b p_T^2)$ over the
kinematic range $p_T^2 < 0.476 x_L^2 \gev^2$.
The solid points are for dijet photoproduction, 
the open points for DIS. 
Statistical uncertainties are shown as vertical bars, where visible. 
The systematic uncertainties, shown as the shaded band,
are similar for both data sets.
}
\label{fig-bxldijdis}
\end{figure}

\Fig{bxldijdis} shows the exponential $p_T^2$ slopes $b(x_L)$
for dijet photoproduction and inclusive DIS.
They are similar in magnitude and both rise with $x_L$.
Although the slopes rise somewhat faster with $x_L$ in
the dijet photoproduction data, 
there is no statistically significant difference between
the two sets except for $x_L>0.9$.

\subsubsection{Comparison of dijet direct and resolved photon contributions}
\label{sec-dirres}

\begin{figure}[t]
\centering
\epsfig{file=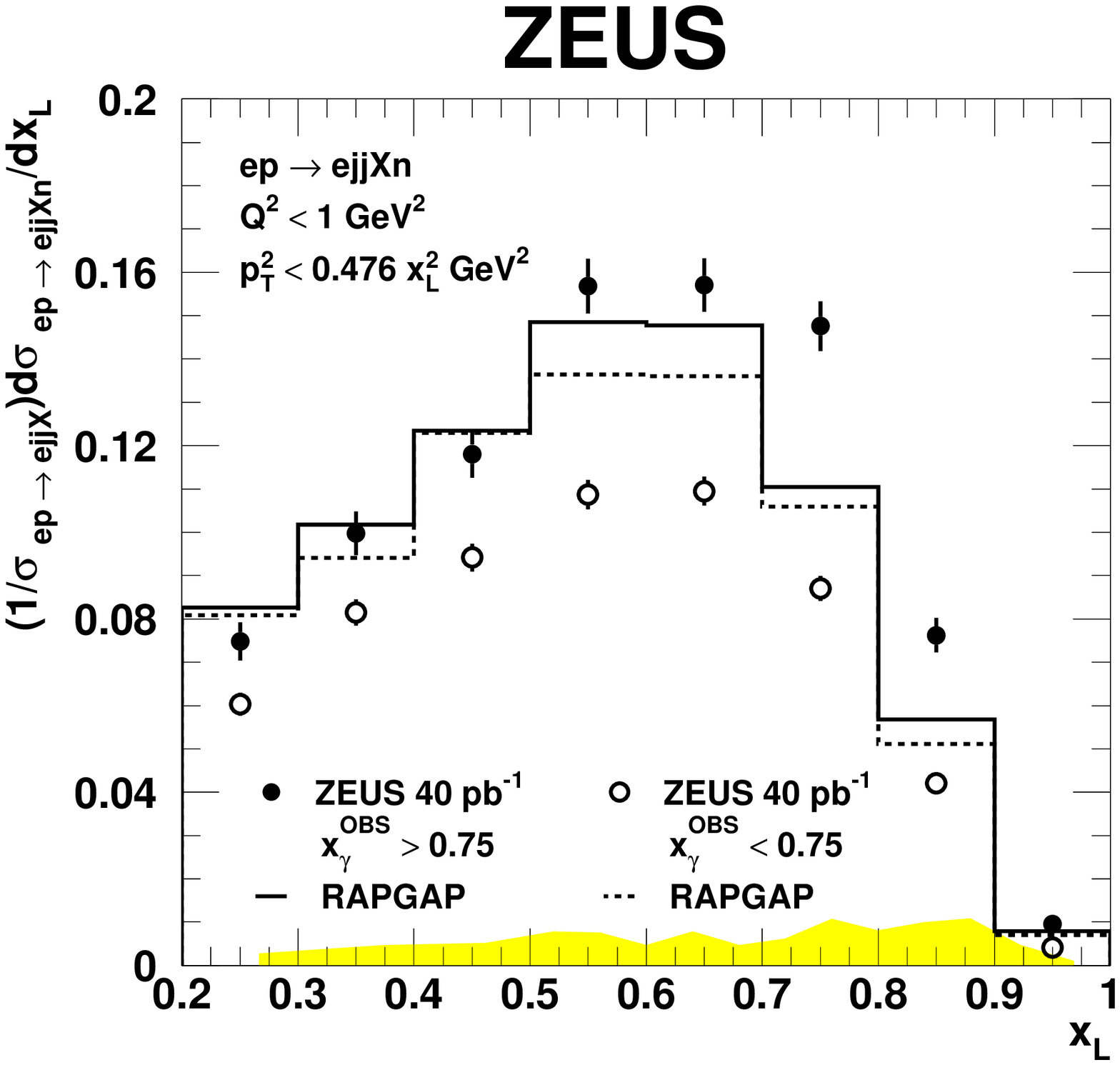,width=14cm}
\caption{
The normalized differential distributions
\mbox{$(1/\sigma_{\inc})d\sigma_{\LN}/dx_L$}
for the direct-enhanced ($x_{\gamma}^{\rm OBS}>0.75$, solid points)
and resolved-enhanced ($x_{\gamma}^{\rm OBS}<0.75$, open points)
dijet photoproduction samples.
Statistical uncertainties are shown as vertical bars. 
The systematic uncertainties, shown as the shaded band,
are similar for both data sets.
The histograms are the predictions of {\sc Rapgap} for
the respective components.
}
\label{fig-xldirres}
\end{figure}

\begin{figure}[t]
\centering
\epsfig{file=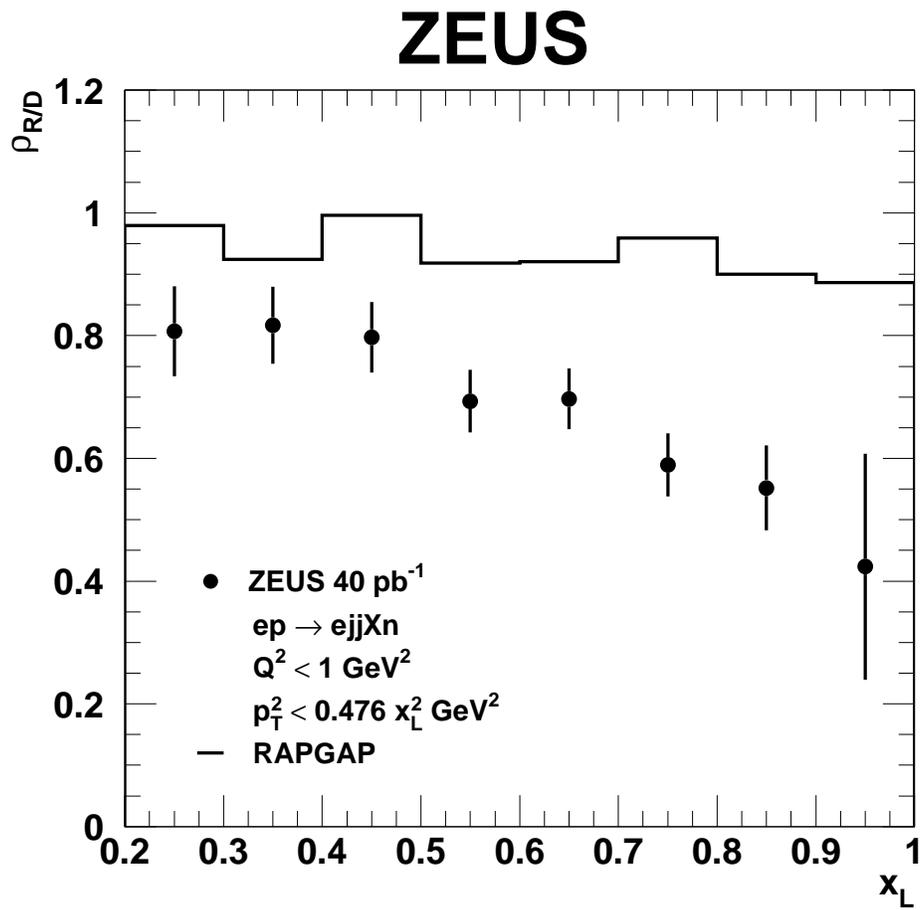,width=14cm}
\caption{ 
Ratio of leading-neutron spectra for resolved ($x_{\gamma}^{\rm OBS}<0.75$)
and direct ($x_{\gamma}^{\rm OBS}>0.75$) contributions to dijet photoproduction.
Only statistical uncertainties are displayed.
The data is compared to the prediction of {\sc Rapgap}.}
\label{fig-ratiodirres}
\end{figure}

The neutron $x_L$ distributions in the dijet photoproduction data,
enriched in direct $(\xgo>0.75)$ and resolved $(\xgo<0.75)$ processes,
are shown in \fig{xldirres},
normalized to their corresponding samples without a neutron requirement.
In the resolved contribution, relatively fewer neutrons are observed.
\Fig{xldirres} also shows the predictions of {\sc Rapgap} for
the $x_L$ distributions of the direct and resolved contributions.
\Fig{ratiodirres} presents the ratio between the resolved and direct
contributions to the cross section,
\begin{equation}
\rho_{\rm \tiny R/D} = \frac{
(1/\sigma_{\inc}) d\sigma_{\LN}/dx_L (\xgo<0.75) }{
(1/\sigma_{\inc}) d\sigma_{\LN}/dx_L (\xgo>0.75) } \, ,
\label{eq-rhord}
\end{equation}
as a function of $x_L$ for data and the {\sc Rapgap} prediction.
The magnitude and shape are not described by {\sc Rapgap}.

\subsection{Role of kinematic constraints}
\label{sec-kinemat}
The $x_L$ distributions for dijet photoproduction and
for DIS are depicted in \fig{xldijdis}.
It is interesting to investigate whether the difference between the
two distributions is a characteristic of the $p\rightarrow n$ transition 
or if it is a kinematic effect, due to different forward energy flows.
To investigate such kinematic constraints, $X_{BP}$,
the fraction of the proton beam energy going into the forward
beampipe region, $\eta \gtrsim 5$, was considered:
\begin{equation}
X_{BP} = 1 - \frac{E+P_Z}{2E_p}.
\label{eq-xbp}
\end{equation}
Here $E_p = 920$ GeV is the proton beam energy
and $E+P_Z$ is the longitudinal energy-momentum,
$E+P_Z = \sum_i E_i (1+\cos{\theta_i})$,
with the sum running over all CAL and FPC cells with
energy $E_i$ and polar angle $\theta_i$.
The energy of the leading neutron in an event is restricted to $x_L<X_{BP}$.

\begin{figure}[t]
\centering
\epsfig{file=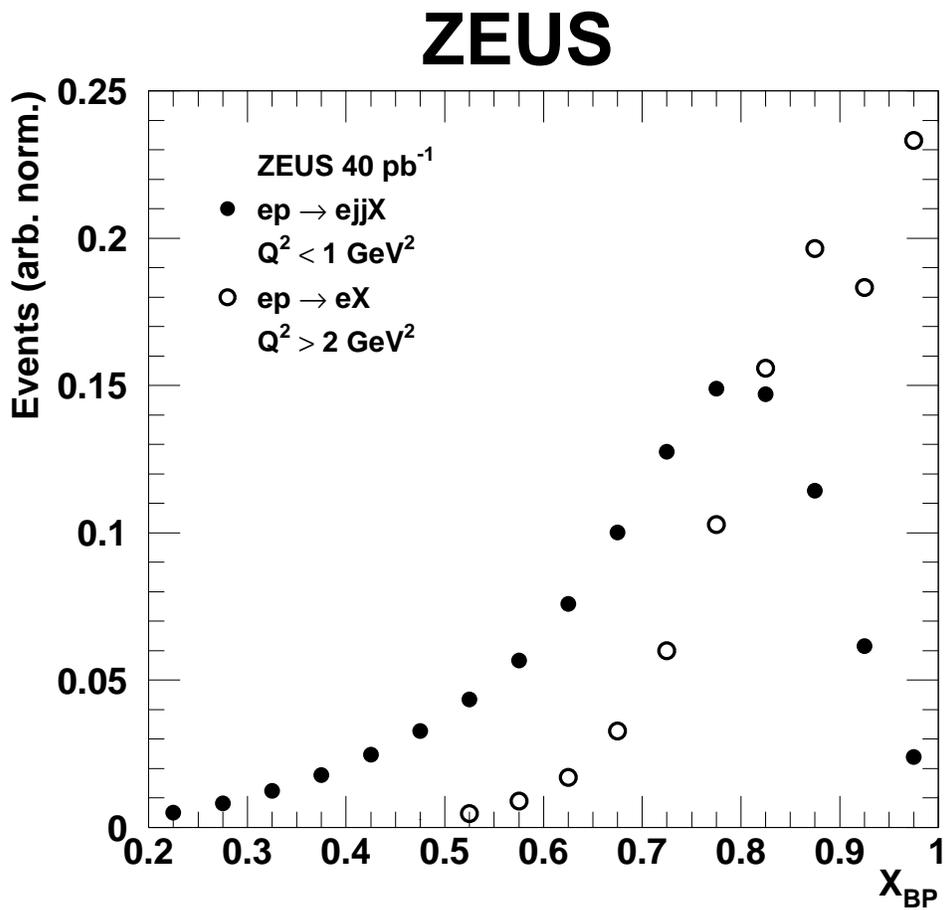,width=14cm}
\caption{
Comparison between $X_{BP}$ distributions for the
dijet photoproduction (solid points) and inclusive DIS (open points)
{\protect \cite{DESY-07-011}} samples.
In both cases no neutron tag was required.
Statistical uncertainties are smaller than the plotted points.
}
\label{fig-xbpdijdis}
\end{figure}

\begin{figure}[t]
\centering
\epsfig{file=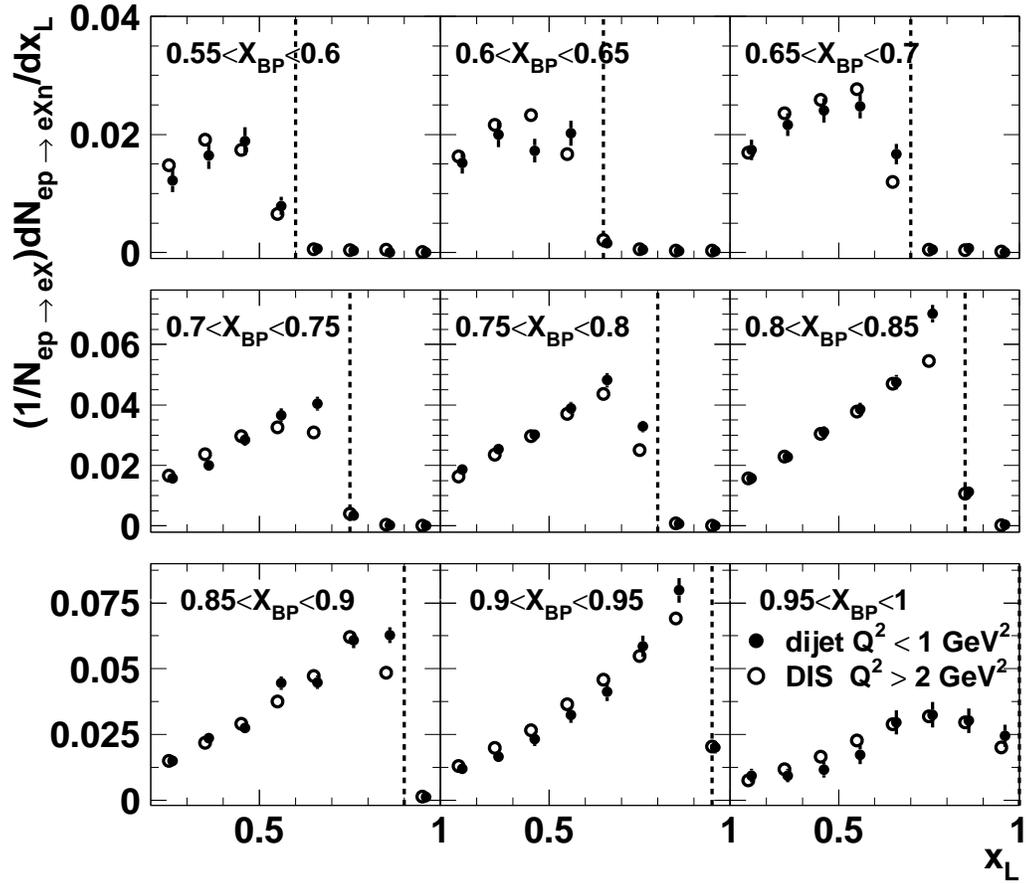,width=14cm}
\caption{
Neutron yield as a function of $x_L$ for different bins of $X_{BP}$ for
the dijet photoproduction (solid points) and inclusive DIS (open points)
samples.
The data are not corrected for detector acceptance. 
Statistical uncertainties are shown as vertical bars, where visible.
The vertical dashed lines show the constraint $x_L<X_{BP}$.
}
\label{fig-xlxbpbins}
\end{figure}

The $X_{BP}$ distributions for dijet photoproduction and
DIS, both without a leading neutron requirement, are shown in
\fig{xbpdijdis}.
The dijet photoproduction data are peaked at significantly
lower $X_{BP}$ and have a much larger tail at very low
$X_{BP}$ than the DIS data.
\Fig{xlxbpbins} shows the neutron $x_L$ 
distributions\footnote{
These $x_L$ distributions are not corrected for acceptance.
The acceptance correction at a given $x_L$ depends only on
the exponential $p_T^2$ slope $b(x_L)$.
As shown in \fig{bxldijdis}, the slopes for dijet photoproduction
and DIS have very similar values.
Differences in the acceptance correction are small and may
be ignored for the comparisons made here.
}
of the dijet photoproduction and DIS data in bins of $X_{BP}$,
normalized by the number of events
without a neutron requirement in the $X_{BP}$ bin.
They reflect the constraint $x_L<X_{BP}$.
For any given value of $X_{BP}$,
the two samples have nearly identical $x_L$ distributions,
both in shape and normalization. This indicates that
a given value of longitudinal energy-momentum measured in
the central detector is associated with the same
neutron yield and spectrum, regardless of whether the process
is dijet photoproduction or DIS.

\begin{figure}[t]
\centering
\epsfig{file=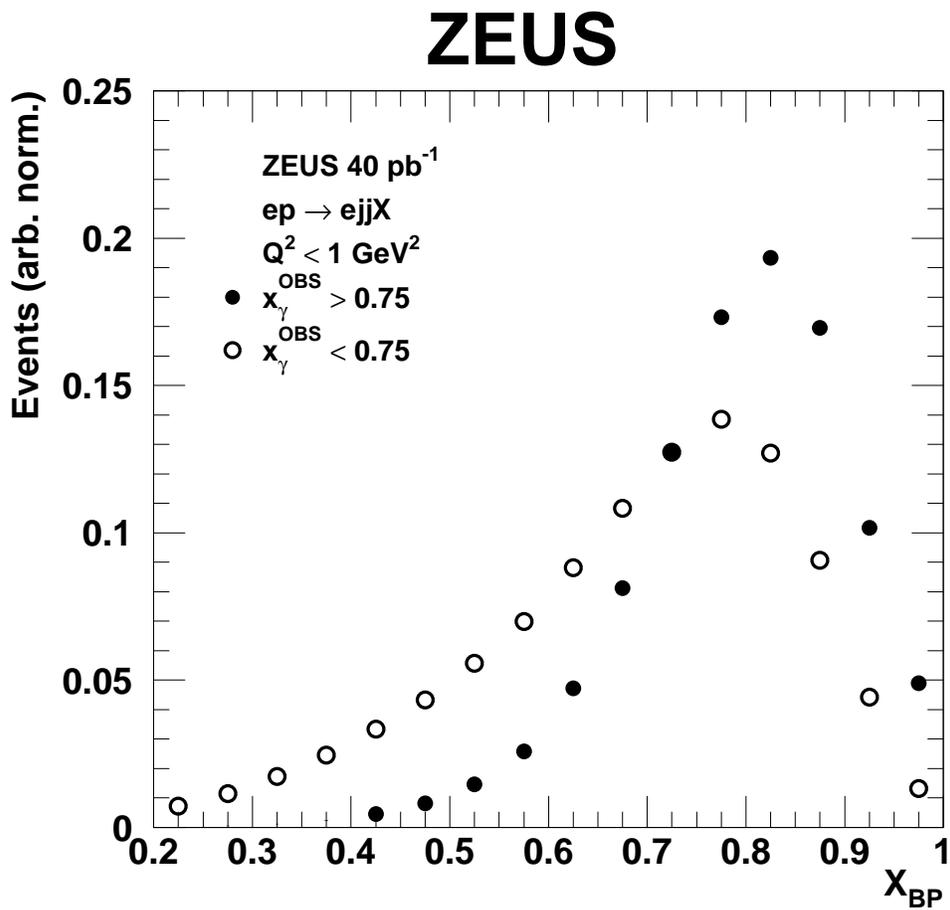,width=14cm}
\caption{
Comparison between $X_{BP}$ distributions for
the dijet photoproduction direct (solid points)
and resolved (open points) photon contributions.
In both cases no neutron tag was required.
Statistical uncertainties are smaller than the plotted points.
}
\label{fig-xbpdirres}
\end{figure}

\begin{figure}[t]
\centering
\epsfig{file=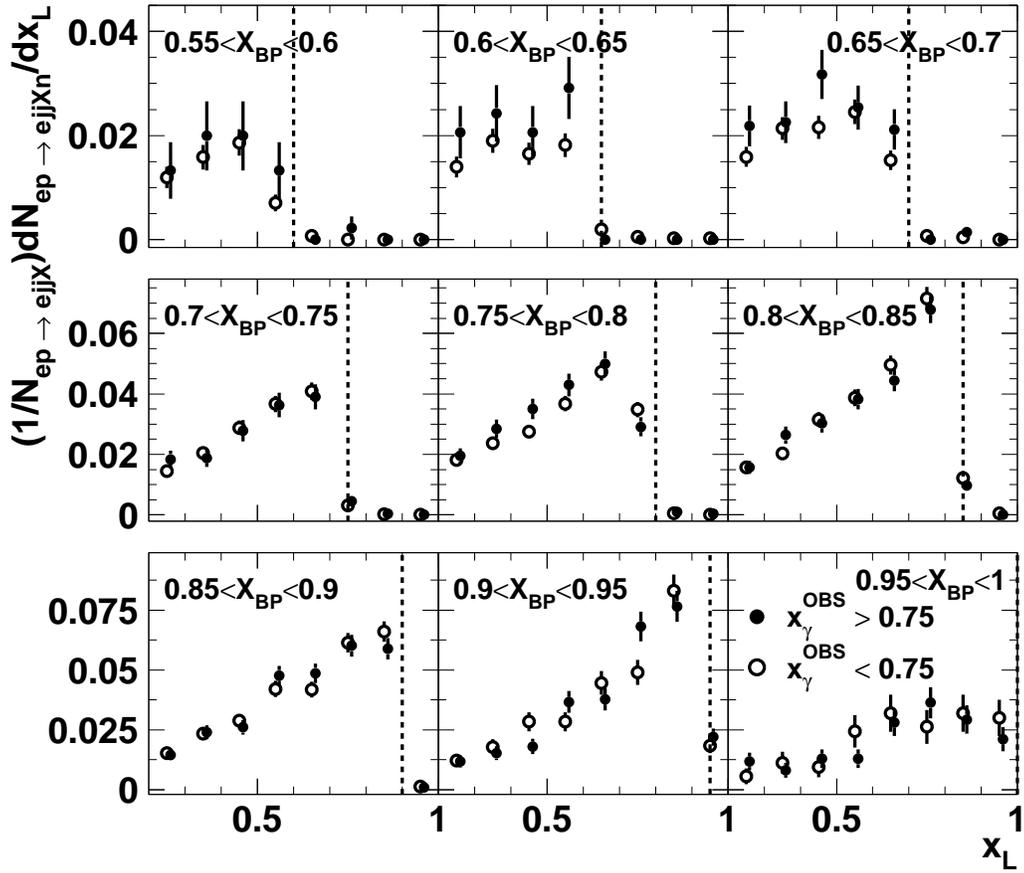,width=14cm}
\caption{
Neutron yield as a function of $x_L$ for different bins of $X_{BP}$ for
the dijet photoproduction direct (solid points)
and resolved (open points) photon contributions.
The data are not corrected for detector acceptance.
Statistical uncertainties are shown as vertical bars, where visible.
The vertical dashed lines show the constraint $x_L<X_{BP}$.
}
\label{fig-xlxbpbinsresdir}
\end{figure}

The effect of kinematic constraints from energy distributions in
the central detector can also be investigated in the
$x_L$ distributions of direct and resolved photoproduction as shown in
\fig{xldirres}.
\Fig{xbpdirres} shows the $X_{BP}$ distributions for the contributions
from direct and resolved photons without a neutron tag being required.
The resolved contribution peaks at lower $X_{BP}$ and has a much
larger tail at very low $X_{BP}$ than the direct contribution.
\Fig{xlxbpbinsresdir} shows the neutron yield as a function of $x_L$
(not corrected for acceptance)
in different bins of $X_{BP}$ for the two contributions.
As in the comparison to DIS, they verify the constraint
$x_L<X_{BP}$, and for any given value of $X_{BP}$,
the two samples have nearly identical $x_L$ distributions,
both in shape and normalization. Thus the neutron $x_L$ spectra
in dijet photoproduction as well as in DIS
seem to depend only on the energy available in the proton-remnant region.

\section{Discussion of rescattering}
\label{sec-discuss}

\begin{figure}[t]
\centering
\epsfig{file=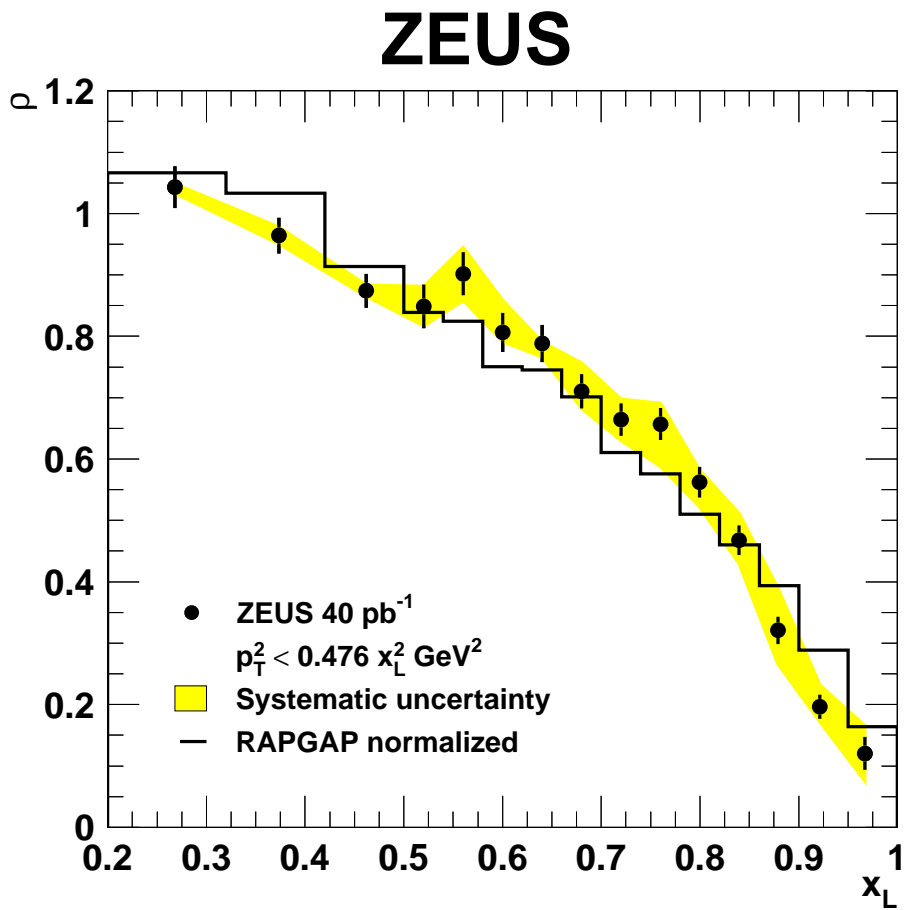,width=14cm}
\caption{
Ratio of the leading-neutron $x_L$ distributions dijet photoproduction to DIS.
The data are the solid points, with statistical uncertainties
shown by vertical bars and the systematic uncertainty
by the shaded band. The histogram is the prediction of {\sc Rapgap}
after its normalization was adjusted to both data sets separately.
}
\label{fig-ratiojjdis}
\end{figure}

\begin{figure}[t]
\centering
\epsfig{file=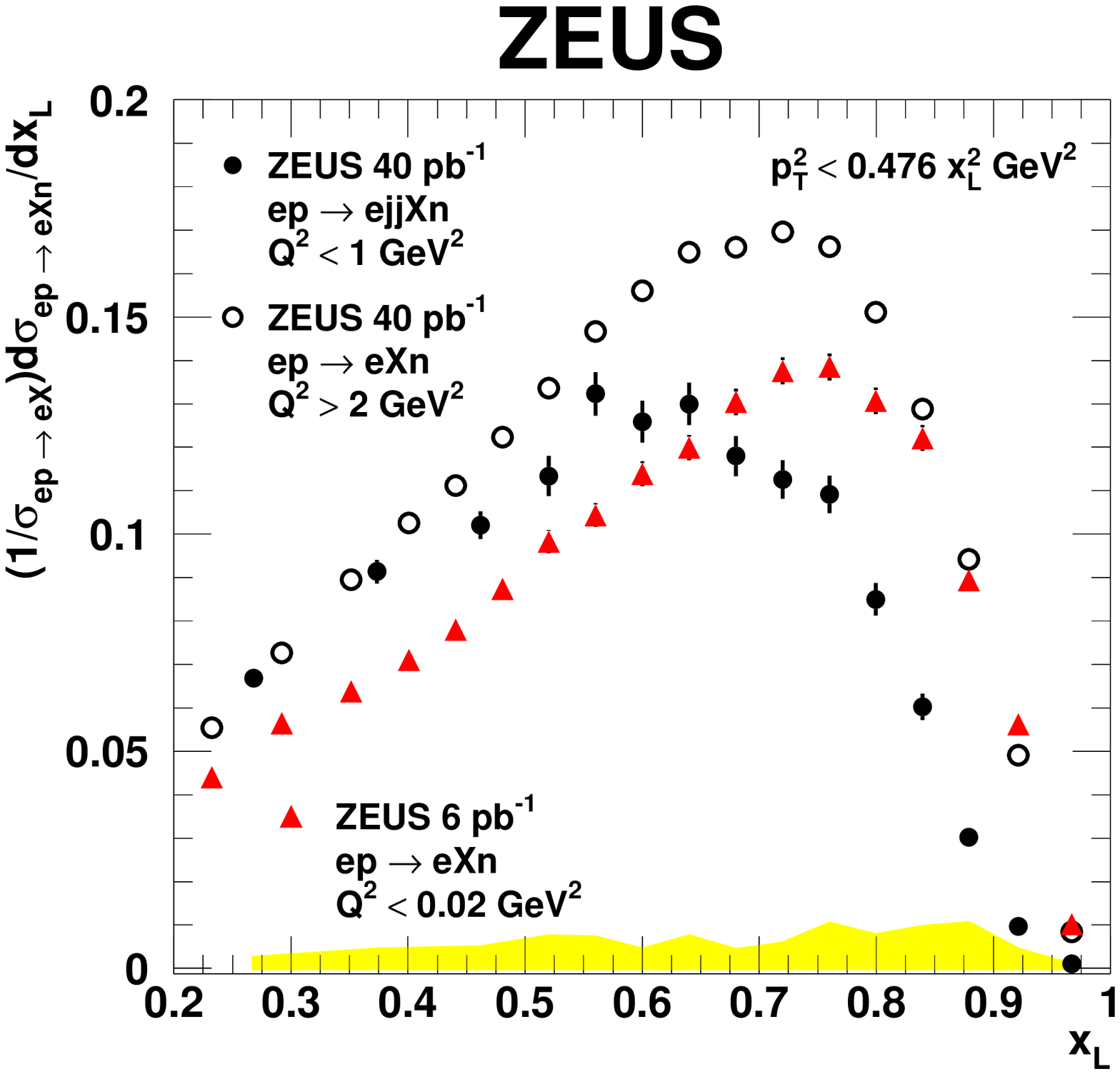,width=14cm}
\caption{
Neutron yields as a function of $x_L$ for
dijet photoproduction (solid points), inclusive DIS (open points),
and inclusive photoproduction (shaded triangles) {\protect \cite{DESY-07-011}}.
Statistical uncertainties are shown as vertical bars, where visible.
The systematic uncertainties,
shown as the shaded band, are similar for all three data sets.
}
\label{fig-xldijdisphp}
\end{figure}

The good statistical accuracy of the data allows an investigation into
effects of rescattering.
The comparison of photoproduction to DIS offers one way to investigate
rescattering effects, which are predicted to result in
a lower neutron yield in photoproduction.
\Fig{xldijdisphp} shows the neutron yield as a function of $x_L$
for dijet photoproduction, inclusive DIS with $Q^2>2\gev^2$,
and inclusive photoproduction $ep \rightarrow eXn$~\cite{DESY-07-011}.
The inclusive photoproduction sample was obtained by tagging
the scattered positron, with a resulting range of $Q^2<0.02\gev^2$.
The neutron yield for the positron-tagged inclusive sample
agrees with the yield observed for inclusive DIS at high values of $x_L$.
At low $x_L$, the neutron yield in inclusive photoproduction is
smaller than in inclusive DIS.
This was shown to be consistent with models of
rescattering~\cite{rescat3,rescat4,epj:c47:385,epj:c48:797}.
The neutron yield is also smaller in dijet photoproduction,
but the $x_L$ dependence of the suppression is reversed.
The neutron yields are similar at low values of $x_L$,
whereas the neutron yield in dijet photoproduction is lower
at high $x_L$ values. This was shown in \fig{ratiojjdis}.

The behavior of the neutron yield for dijet photoproduction
is inconsistent with the rescattering models that
described the yield for the positron-tagged photoproduction sample.
Information concerning rescattering might be difficult to obtain from a direct
comparison of dijet photoproduction and inclusive DIS data because of the
different hadronic final states.
A Regge factorization model without rescattering effects ({\sc Rapgap})
can reproduce reasonably the differences in neutron yields between
dijet photoproduction and inclusive DIS.
A qualitative explanation can be deduced from Eq.~\eq{opef}: The cross section
is proportional to $\sigma_{e\pi \rightarrow eX}(s^\prime)$,
and this cross section rises steeply with $s^\prime = (1-x_L)s$
for dijet production~\cite{pl:b560:7},
whereas the cross section for the inclusive DIS reaction depends
only weakly on $s^\prime$~\cite{epj:c7:609,np:b713:3}.
Therefore one expects a drop of the ratio of the
dijet to DIS neutron yields as $s^\prime \propto (1-x_L)$ goes to 0. 
This is seen in \fig{ratiojjdis}.

Another way to look for such effects is to compare
the direct and resolved contributions to dijet photoproduction.
For the direct photon contribution, the photon is assumed to be pointlike;
for the resolved photon interactions, the photon is assumed
to have size and structure.
This structure may be expected to increase the probability
of rescattering.
Indeed, the lower neutron yield in the resolved contribution
to the cross section, as shown in the $\xgo$ distribution in \fig{xgamwxp}, 
seems to indicate such a loss mechanism.
However, this seems in contradiction with the
$x_L$ dependence of the effect, as shown in \figand{xldirres}{ratiodirres}.
These figures show that the 
neutron yield in the resolved contribution decreases relative
to the yield in the direct contribution for increasing values of $x_L$.
This contradicts the predictions from the rescattering models
which described the behavior of the inclusive photoproduction
sample~\cite{DESY-07-011}, where the effect goes in the opposite direction.
Again, a comparison is complicated by the different hadronic final
states in the direct and resolved contributions.

In summary, no clear conclusion on the presence of rescattering
effects in dijet photoproduction can be drawn from the data alone.
Only a comparison to a specific model could clarify this issue.

\section{Summary}

Differential cross sections for neutron-tagged and untagged
dijet photoproduction,
$e^+ + p \rightarrow e^+ + \mathrm{jet} + \mathrm{jet} + X \, ( + n)$,
have been measured.
The measurements required jets with
$\ETJa>7.5$~GeV,
$\ETJb>6.5$~GeV and
\mbox{$-1.5<\ETAJ<2.5$},
in the kinematic region \mbox{$Q^2<1$~GeV$^2$} and 
\mbox{$130<W<280$ GeV}, with the additional restriction of
\mbox{$x_L>0.2$} and \mbox{$\theta_n<0.75$ mrad} on the neutron-tagged
sample.
The cross sections were measured as functions of 
$\ETJ$, $\ETAJ$, $W$, $\xgo$ and $\xpo$.

The ratios of the neutron-tagged to untagged differential cross sections show
a reduction of the neutron yield at low $\xgo$, large $\ETAJ$,
and large $\xpo$.
These regions are dominated by resolved photon events.

The normalized leading-neutron $x_L$ distribution was measured. It is in
reasonable agreement with the {\sc Rapgap} MC model including pion
exchange, which is essential to obtain a reasonable description of the data.
In addition, the leading-neutron
cross section as a function of $\xpio$, the 
fraction of the exchanged pion four-momentum entering the hard scattering, was 
measured in the restricted kinematic range $x_L>0.6$, where pion exchange is 
the dominant production process, and good agreement with the model was found.

The leading-neutron cross sections as a function of $p_T^2$ in
different regions of $x_L$ were measured in dijet photoproduction.
The $p_T^2$ distributions are well described by exponentials,
and the two-dimensional $(x_L, p_T^2)$ distribution is fully characterized by
the slopes and intercepts from exponential fits in each $x_L$ bin.

The relation between the neutron yield and the fraction of the proton
beam energy going into the forward beam pipe region, $X_{BP}$, was studied.
The relative neutron rate as a function of $x_L$
seems to depend only on $X_{BP}$.
This effect accounts for the observed differences between the
$x_L$ distributions of the photoproduction dijet and the DIS data samples,
and between those of the direct and resolved dijet samples.

No clear conclusion on the presence of rescattering effects can be
drawn.  While the reduction of the neutron yield in the region enriched
in resolved photons is suggestive of the presence of a rescattering
effect, the fact that this yield reduction is mainly at large
$x_L$ seems to contradict the basic expectations of rescattering models.

\section*{Acknowledgments}

We appreciate the contributions to the construction and maintenance
of the ZEUS detector of many people who are not listed as authors.
The HERA machine group and the DESY computing staff
are especially acknowledged for their success
in providing excellent operation of the collider
and the data-analysis environment.
We thank the DESY directorate for their strong support and encouragement, 
and are especially grateful for their financial support which made
possible the construction and installation of the FNC.
We are also happy to acknowledge the DESY accelerator group
for allowing the installation of the FNC in close
proximity to the HERA machine components.

\vfill\eject


\providecommand{\etal}{et al.\xspace}
\providecommand{\coll}{Coll.\xspace}
\catcode`\@=11
\def\@bibitem#1{%
\ifmc@bstsupport
  \mc@iftail{#1}%
    {;\newline\ignorespaces}%
    {\ifmc@first\else.\fi\orig@bibitem{#1}}
  \mc@firstfalse
\else
  \mc@iftail{#1}%
    {\ignorespaces}%
    {\orig@bibitem{#1}}%
\fi}%
\catcode`\@=12



\begin{table}[p]
\begin{center}

\begin{tabular}{|c|c|c|c|}
\hline
$\ETJ$ (GeV) & $d\sigma/d\ETJ$ (nb/GeV) & $d\sigma_{\rm LN}/d\ETJ$ (nb/GeV) & $r_{\rm LN}$ (\%)  \\
\hline
  $\;\;7.6$ & $ 8.414 \pm 0.021^{+0.839}_{-0.842}\,^{+0.430}_{-0.437} $ &
$ 0.572 \pm 0.007^{+0.056}_{-0.057}\,^{+0.037}_{-0.035} $ & $  6.80 \pm  0.09^{+ 0.19}_{- 0.20}\,^{+ 0.32}_{- 0.10} $ \\
  $\;\;9.7$ & $ 5.299 \pm 0.015^{+0.388}_{-0.391}\,^{+0.254}_{-0.171} $ &
$ 0.368 \pm 0.005^{+0.027}_{-0.027}\,^{+0.021}_{-0.011} $ & $  6.95 \pm  0.10^{+ 0.30}_{- 0.30}\,^{+ 0.09}_{- 0.03} $ \\
 11.9 & $ 2.336 \pm 0.008^{+0.062}_{-0.062}\,^{+0.106}_{-0.189} $ &
$ 0.162 \pm 0.003^{+0.005}_{-0.004}\,^{+0.009}_{-0.015} $ & $  6.93 \pm  0.12^{+ 0.19}_{- 0.16}\,^{+ 0.09}_{- 0.09} $ \\
 14.0 & $ 1.0538 \pm 0.0054^{+0.0253}_{-0.0247}\,^{+0.1119}_{-0.1065} $ &
$ 0.0737 \pm 0.0018^{+0.0030}_{-0.0017}\,^{+0.0061}_{-0.0071} $ & $  6.99 \pm  0.17^{+ 0.20}_{- 0.20}\,^{+ 0.12}_{- 0.17} $ \\
 16.2 & $ 0.5188 \pm 0.0038^{+0.0201}_{-0.0213}\,^{+0.0561}_{-0.0557} $ &
$ 0.0383 \pm 0.0013^{+0.0027}_{-0.0015}\,^{+0.0032}_{-0.0043} $ & $  7.38 \pm  0.25^{+ 0.24}_{- 0.24}\,^{+ 0.12}_{- 0.14} $ \\
 19.4 & $ 0.2094 \pm 0.0017^{+0.0042}_{-0.0063}\,^{+0.0255}_{-0.0233} $ &
$ 0.0157 \pm 0.0006^{+0.0003}_{-0.0003}\,^{+0.0018}_{-0.0017} $ & $  7.49 \pm  0.29^{+ 0.17}_{- 0.17}\,^{+ 0.06}_{- 0.11} $ \\
 23.6 & $ 0.0686 \pm 0.0010^{+0.0024}_{-0.0024}\,^{+0.0090}_{-0.0075} $ &
$ 0.0046 \pm 0.0003^{+0.0002}_{-0.0002}\,^{+0.0007}_{-0.0005} $ & $  6.69 \pm  0.48^{+ 0.29}_{- 0.29}\,^{+ 0.10}_{- 0.05} $ \\
 27.9 & $ 0.0255 \pm 0.0006^{+0.0015}_{-0.0020}\,^{+0.0029}_{-0.0033} $ &
$ 0.0015 \pm 0.0002^{+0.0001}_{-0.0002}\,^{+0.0002}_{-0.0002} $ & $  5.76 \pm  0.73^{+ 0.20}_{- 0.20}\,^{+ 0.05}_{- 0.05} $ \\
\hline
\end{tabular}

\vspace{0.5cm}

\begin{tabular}{|c|c|c|c|}
\hline
$\ETAJ$ & $d\sigma/d\ETAJ$ (nb) & $d\sigma_{\rm LN}/d\ETAJ$ (nb) & $r_{\rm LN}$ (\%)  \\
\hline
 $-1.33$ & $  0.80 \pm  0.01^{+ 0.13}_{- 0.15}\,^{+ 0.10}_{- 0.11} $ &
$ 0.078 \pm 0.006^{+0.014}_{-0.013}\,^{+0.010}_{-0.011} $ & $  9.69 \pm  0.73^{+ 1.66}_{- 0.67}\,^{+ 0.18}_{- 0.21} $ \\
 $-1.00$ & $  3.07 \pm  0.03^{+ 0.17}_{- 0.27}\,^{+ 0.33}_{- 0.37} $ &
$ 0.259 \pm 0.012^{+0.016}_{-0.011}\,^{+0.012}_{-0.012} $ & $  8.42 \pm  0.41^{+ 0.42}_{- 0.42}\,^{+ 0.48}_{- 0.98} $ \\
 $-0.67$ & $  6.10 \pm  0.04^{+ 0.41}_{- 0.51}\,^{+ 0.70}_{- 0.64} $ &
$ 0.478 \pm 0.017^{+0.031}_{-0.028}\,^{+0.070}_{-0.024} $ & $  7.84 \pm  0.28^{+ 0.41}_{- 0.13}\,^{+ 0.52}_{- 0.52} $ \\
 $-0.33$ & $  8.84 \pm  0.05^{+ 0.79}_{- 0.86}\,^{+ 0.77}_{- 0.91} $ &
$ 0.744 \pm 0.019^{+0.068}_{-0.067}\,^{+0.044}_{-0.057} $ & $  8.41 \pm  0.22^{+ 0.19}_{- 0.19}\,^{+ 0.32}_{- 0.82} $ \\
  $\;\;\;0.00$ & $ 11.00 \pm  0.05^{+ 0.58}_{- 0.62}\,^{+ 0.79}_{- 0.82} $ &
$ 0.847 \pm 0.019^{+0.047}_{-0.044}\,^{+0.057}_{-0.042} $ & $  7.70 \pm  0.18^{+ 0.15}_{- 0.15}\,^{+ 0.45}_{- 0.31} $ \\
  $\;\;\;0.33$ & $ 12.43 \pm  0.05^{+ 0.46}_{- 0.47}\,^{+ 0.75}_{- 0.81} $ &
$ 0.926 \pm 0.020^{+0.036}_{-0.036}\,^{+0.071}_{-0.054} $ & $  7.45 \pm  0.16^{+ 0.17}_{- 0.17}\,^{+ 0.49}_{- 0.14} $ \\
  $\;\;\;0.67$ & $ 12.98 \pm  0.06^{+ 0.46}_{- 0.46}\,^{+ 0.81}_{- 0.87} $ &
$ 0.897 \pm 0.019^{+0.031}_{-0.040}\,^{+0.075}_{-0.062} $ & $  6.91 \pm  0.15^{+ 0.15}_{- 0.21}\,^{+ 0.44}_{- 0.10} $ \\
  $\;\;\;1.00$ & $ 12.46 \pm  0.06^{+ 0.55}_{- 0.56}\,^{+ 0.81}_{- 0.81} $ &
$ 0.820 \pm 0.020^{+0.035}_{-0.035}\,^{+0.046}_{-0.067} $ & $  6.58 \pm  0.16^{+ 0.13}_{- 0.18}\,^{+ 0.19}_{- 0.14} $ \\
  $\;\;\;1.33$ & $ 11.84 \pm  0.06^{+ 0.87}_{- 0.86}\,^{+ 0.85}_{- 0.80} $ &
$ 0.779 \pm 0.019^{+0.053}_{-0.053}\,^{+0.076}_{-0.070} $ & $  6.58 \pm  0.16^{+ 0.12}_{- 0.19}\,^{+ 0.26}_{- 0.32} $ \\
  $\;\;\;1.67$ & $ 12.44 \pm  0.06^{+ 1.42}_{- 1.41}\,^{+ 1.06}_{- 1.03} $ &
$ 0.735 \pm 0.017^{+0.080}_{-0.081}\,^{+0.085}_{-0.066} $ & $  5.91 \pm  0.14^{+ 0.13}_{- 0.14}\,^{+ 0.29}_{- 0.11} $ \\
  $\;\;\;2.00$ & $ 13.53 \pm  0.06^{+ 1.83}_{- 1.82}\,^{+ 1.21}_{- 1.25} $ &
$ 0.800 \pm 0.019^{+0.108}_{-0.107}\,^{+0.112}_{-0.098} $ & $  5.91 \pm  0.14^{+ 0.07}_{- 0.07}\,^{+ 0.38}_{- 0.19} $ \\
  $\;\;\;2.33$ & $ 12.76 \pm  0.06^{+ 1.18}_{- 1.13}\,^{+ 1.07}_{- 1.07} $ &
$ 0.707 \pm 0.018^{+0.067}_{-0.064}\,^{+0.130}_{-0.081} $ & $  5.54 \pm  0.14^{+ 0.04}_{- 0.04}\,^{+ 0.62}_{- 0.26} $ \\
\hline
\end{tabular}

\caption{
Differential cross-sections $\sigma_{\rm (LN)}$ for the processes
$ e^+ + p \rightarrow e^+ + \mathrm{jet} + \mathrm{jet}+ X ( + n )$
and the ratio $r_{\rm LN} = \sigma_{\rm LN}/\sigma$ as functions of
$E_T$ and $\eta$.
For each cross section and ratio, the first uncertainty is statistical,
the second systematic, excluding the CAL energy scale,
and the third the systematic due to the CAL energy scale.
}
\label{tab-eteta}
\end{center}
\end{table}

\clearpage


\begin{table}[p]
\begin{center}

\begin{tabular}{|c|c|c|c|}
\hline
$\xgo$ & $d\sigma/d\xgo$ (nb) & $d\sigma_{\rm LN}/d\xgo$ (nb) & $r_{\rm LN}$ (\%)  \\
\hline
 0.07 & $ 15.54 \pm  0.11^{+ 2.84}_{- 2.48}\,^{+ 1.49}_{- 2.07} $ &
$ 0.586 \pm 0.024^{+0.103}_{-0.088}\,^{+0.122}_{-0.111} $ & $  3.77 \pm  0.16^{+ 0.12}_{- 0.05}\,^{+ 0.51}_{- 0.29} $ \\
 0.21 & $ 23.22 \pm  0.12^{+ 2.56}_{- 2.28}\,^{+ 1.88}_{- 2.20} $ &
$ 1.174 \pm 0.034^{+0.137}_{-0.110}\,^{+0.162}_{-0.148} $ & $  5.06 \pm  0.15^{+ 0.08}_{- 0.08}\,^{+ 0.41}_{- 0.17} $ \\
 0.36 & $ 17.13 \pm  0.10^{+ 1.60}_{- 1.56}\,^{+ 1.18}_{- 1.24} $ &
$ 1.017 \pm 0.031^{+0.088}_{-0.086}\,^{+0.102}_{-0.068} $ & $  5.94 \pm  0.18^{+ 0.17}_{- 0.16}\,^{+ 0.39}_{- 0.19} $ \\
 0.50 & $ 14.92 \pm  0.09^{+ 1.31}_{- 1.32}\,^{+ 1.03}_{- 1.04} $ &
$ 1.060 \pm 0.032^{+0.091}_{-0.098}\,^{+0.052}_{-0.079} $ & $  7.10 \pm  0.22^{+ 0.21}_{- 0.21}\,^{+ 0.30}_{- 0.54} $ \\
 0.64 & $ 17.09 \pm  0.10^{+ 2.18}_{- 2.18}\,^{+ 1.17}_{- 1.21} $ &
$ 1.283 \pm 0.037^{+0.164}_{-0.165}\,^{+0.075}_{-0.084} $ & $  7.51 \pm  0.22^{+ 0.22}_{- 0.22}\,^{+ 0.56}_{- 0.32} $ \\
 0.79 & $ 28.39 \pm  0.13^{+ 1.76}_{- 1.77}\,^{+ 1.60}_{- 1.84} $ &
$ 2.317 \pm 0.052^{+0.151}_{-0.138}\,^{+0.148}_{-0.174} $ & $  8.16 \pm  0.19^{+ 0.04}_{- 0.04}\,^{+ 0.36}_{- 0.45} $ \\
 0.93 & $ 21.35 \pm  0.11^{+ 2.16}_{- 2.07}\,^{+ 0.61}_{- 1.00} $ &
$ 1.949 \pm 0.046^{+0.158}_{-0.132}\,^{+0.069}_{-0.085} $ & $  9.13 \pm  0.22^{+ 0.34}_{- 0.20}\,^{+ 0.24}_{- 0.11} $ \\
\hline
\end{tabular}

\vspace{0.5cm}

\begin{tabular}{|c|c|c|c|}
\hline
$W$ (GeV) & $d\sigma/dW$ (nb/GeV) & $d\sigma_{\rm LN}/dW$ (nb/GeV) & $r_{\rm LN}$ (\%)  \\
\hline
 142 & $ 0.109 \pm 0.001^{+0.008}_{-0.008}\,^{+0.009}_{-0.005} $ &
$ 0.0090 \pm 0.0003^{+0.0006}_{-0.0007}\,^{+0.0012}_{-0.0006} $ & $  8.30 \pm  0.24^{+ 0.07}_{- 0.07}\,^{+ 0.43}_{- 0.25} $ \\
 167 & $ 0.137 \pm 0.001^{+0.011}_{-0.011}\,^{+0.012}_{-0.010} $ &
$ 0.0097 \pm 0.0002^{+0.0008}_{-0.0008}\,^{+0.0013}_{-0.0009} $ & $  7.10 \pm  0.18^{+ 0.08}_{- 0.07}\,^{+ 0.28}_{- 0.12} $ \\
 192 & $ 0.143 \pm 0.001^{+0.009}_{-0.009}\,^{+0.011}_{-0.011} $ &
$ 0.0095 \pm 0.0002^{+0.0006}_{-0.0006}\,^{+0.0011}_{-0.0009} $ & $  6.62 \pm  0.17^{+ 0.01}_{- 0.01}\,^{+ 0.37}_{- 0.17} $ \\
 217 & $ 0.140 \pm 0.001^{+0.013}_{-0.013}\,^{+0.010}_{-0.009} $ &
$ 0.0089 \pm 0.0002^{+0.0008}_{-0.0008}\,^{+0.0009}_{-0.0007} $ & $  6.35 \pm  0.17^{+ 0.14}_{- 0.14}\,^{+ 0.36}_{- 0.29} $ \\
 242 & $ 0.132 \pm 0.001^{+0.007}_{-0.007}\,^{+0.012}_{-0.013} $ &
$ 0.0088 \pm 0.0002^{+0.0005}_{-0.0005}\,^{+0.0007}_{-0.0007} $ & $  6.66 \pm  0.17^{+ 0.21}_{- 0.21}\,^{+ 0.21}_{- 0.06} $ \\
 267 & $ 0.127 \pm 0.001^{+0.007}_{-0.007}\,^{+0.014}_{-0.016} $ &
$ 0.0078 \pm 0.0002^{+0.0004}_{-0.0004}\,^{+0.0007}_{-0.0007} $ & $  6.16 \pm  0.17^{+ 0.12}_{- 0.12}\,^{+ 0.31}_{- 0.32} $ \\
\hline
\end{tabular}

\vspace{0.5cm}

\begin{tabular}{|c|c|c|c|}
\hline
$\log_{10}(\xpo)$ & $d\sigma/d\log_{10}(\xpo)$ (nb) & $d\sigma_{\rm LN}/d\log_{10}(\xpo)$ (nb) & $r_{\rm LN}$ (\%)  \\
\hline
 $-2.3$ & $  1.47 \pm  0.02^{+ 0.15}_{- 0.25}\,^{+ 0.62}_{- 0.39} $ &
$ 0.154 \pm 0.009^{+0.014}_{-0.018}\,^{+0.065}_{-0.023} $ & $ 10.49 \pm  0.62^{+ 1.75}_{- 1.62}\,^{+ 0.91}_{- 0.86} $ \\
 $-2.1$ & $  5.33 \pm  0.04^{+ 0.25}_{- 0.42}\,^{+ 1.24}_{- 0.66} $ &
$ 0.488 \pm 0.016^{+0.017}_{-0.023}\,^{+0.077}_{-0.041} $ & $  9.16 \pm  0.31^{+ 0.54}_{- 0.44}\,^{+ 0.44}_{- 0.57} $ \\
 $-1.9$ & $ 11.29 \pm  0.06^{+ 0.51}_{- 0.57}\,^{+ 0.77}_{- 0.68} $ &
$ 0.861 \pm 0.023^{+0.049}_{-0.039}\,^{+0.065}_{-0.056} $ & $  7.63 \pm  0.21^{+ 0.17}_{- 0.17}\,^{+ 0.65}_{- 0.54} $ \\
 $-1.7$ & $ 16.59 \pm  0.08^{+ 0.45}_{- 0.46}\,^{+ 0.99}_{- 1.02} $ &
$ 1.262 \pm 0.030^{+0.034}_{-0.034}\,^{+0.109}_{-0.069} $ & $  7.61 \pm  0.19^{+ 0.20}_{- 0.24}\,^{+ 0.36}_{- 0.34} $ \\
 $-1.5$ & $ 21.43 \pm  0.10^{+ 1.88}_{- 1.92}\,^{+ 1.75}_{- 1.52} $ &
$ 1.475 \pm 0.034^{+0.127}_{-0.129}\,^{+0.231}_{-0.161} $ & $  6.88 \pm  0.16^{+ 0.28}_{- 0.28}\,^{+ 0.60}_{- 0.37} $ \\
 $-1.3$ & $ 24.51 \pm  0.11^{+ 2.27}_{- 2.23}\,^{+ 2.92}_{- 2.13} $ &
$ 1.558 \pm 0.037^{+0.136}_{-0.136}\,^{+0.258}_{-0.171} $ & $  6.35 \pm  0.15^{+ 0.19}_{- 0.20}\,^{+ 0.33}_{- 0.21} $ \\
 $-1.1$ & $ 14.43 \pm  0.09^{+ 1.06}_{- 1.02}\,^{+ 1.62}_{- 1.87} $ &
$ 0.766 \pm 0.027^{+0.053}_{-0.055}\,^{+0.074}_{-0.090} $ & $  5.31 \pm  0.19^{+ 0.06}_{- 0.06}\,^{+ 0.16}_{- 0.09} $ \\
 $-0.9$ & $  3.16 \pm  0.04^{+ 0.29}_{- 0.29}\,^{+ 0.35}_{- 0.35} $ &
$ 0.145 \pm 0.011^{+0.016}_{-0.014}\,^{+0.017}_{-0.015} $ & $  4.61 \pm  0.37^{+ 0.21}_{- 0.09}\,^{+ 0.14}_{- 0.11} $ \\
 $-0.7$ & $  0.38 \pm  0.01^{+ 0.03}_{- 0.03}\,^{+ 0.05}_{- 0.05} $ &
$ 0.018 \pm 0.004^{+0.002}_{-0.002}\,^{+0.001}_{-0.001} $ & $  4.87 \pm  1.09^{+ 0.35}_{- 0.35}\,^{+ 0.14}_{- 0.77} $ \\
\hline
\end{tabular}

\caption{
Differential cross-sections $\sigma_{\rm (LN)}$ for the processes
$ e^+ + p \rightarrow e^+ + \mathrm{jet} + \mathrm{jet}+ X ( + n )$
and the ratio $r_{\rm LN} = \sigma_{\rm LN}/\sigma$ as functions of
$\xgo$, $W$ and $\eta$.
Details are as in  Table \ref{tab-eteta}.
}
\label{tab-xgamwxp}
\end{center}
\end{table}

\clearpage


\begin{table}[p]
\begin{center}

\begin{tabular}{|c|c|}
\hline
$\log_{10}(\xpio)$ & $d\sigma_{\rm LN}/d\log_{10}(\xpio) (\nb)$ \\
\hline
 $-2.1$ & $ 0.0010 \pm 0.0004^{+0.0005}_{-0.0008}\,^{+0.0008}_{-0.0008} $ \\
 $-1.9$ & $ 0.0226 \pm 0.0032^{+0.0053}_{-0.0053}\,^{+0.0074}_{-0.0076} $ \\
 $-1.7$ & $ 0.111 \pm 0.008^{+0.009}_{-0.017}\,^{+0.020}_{-0.015} $ \\
 $-1.5$ & $ 0.248 \pm 0.012^{+0.007}_{-0.009}\,^{+0.017}_{-0.017} $ \\
 $-1.3$ & $ 0.432 \pm 0.017^{+0.015}_{-0.009}\,^{+0.037}_{-0.029} $ \\
 $-1.1$ & $ 0.561 \pm 0.020^{+0.054}_{-0.054}\,^{+0.070}_{-0.033} $ \\
 $-0.9$ & $ 0.653 \pm 0.022^{+0.036}_{-0.036}\,^{+0.082}_{-0.088} $ \\
 $-0.7$ & $ 0.550 \pm 0.021^{+0.060}_{-0.055}\,^{+0.130}_{-0.051} $ \\
 $-0.5$ & $ 0.258 \pm 0.014^{+0.023}_{-0.020}\,^{+0.040}_{-0.037} $ \\
 $-0.3$ & $ 0.0664 \pm 0.0068^{+0.0058}_{-0.0058}\,^{+0.0108}_{-0.0108} $ \\
 $-0.1$ & $ 0.0070 \pm 0.0021^{+0.0010}_{-0.0008}\,^{+0.0020}_{-0.0022} $ \\
\hline
\end{tabular}

\caption{
Differential cross-section $d\sigma_{\rm LN}/d\log_{10}(\xpio)$ for the processes
$ e^+ + p \rightarrow e^+ + \mathrm{jet} + \mathrm{jet}+ X  + n $ for $x_L>0.6$.
Details are as in  Table \ref{tab-eteta}.
}
\label{tab-xpion}
\end{center}
\end{table}

\clearpage


\begin{table}[p]
\begin{center}
{\scriptsize
\begin{tabular}{|c|c|c|c|}
\hline
$x_L$ range & $\langle x_L \rangle$ & $p_T^2(\gev^2)$ &
$\sigma_{\rm norm.} \: (\gev^{-2})$ \\
\hline
 0.20-0.50 & 0.38 & 7.74 $\cdot 10^{-4}$ & 1.797 $\pm$ 0.169 \\
           &      & 2.52 $\cdot 10^{-3}$ & 1.659 $\pm$ 0.156 \\
           &      & 4.86 $\cdot 10^{-3}$ & 1.699 $\pm$ 0.155 \\
           &      & 7.97 $\cdot 10^{-3}$ & 1.511 $\pm$ 0.151 \\
           &      & 1.18 $\cdot 10^{-2}$ & 1.492 $\pm$ 0.149 \\
           &      & 1.65 $\cdot 10^{-2}$ & 1.585 $\pm$ 0.149 \\
 \hline
 0.50-0.58 & 0.54 & 4.84 $\cdot 10^{-3}$ & 1.135 $\pm$ 0.092 \\
           &      & 1.58 $\cdot 10^{-2}$ & 1.008 $\pm$ 0.107 \\
           &      & 3.03 $\cdot 10^{-2}$ & 0.808 $\pm$ 0.095 \\
           &      & 4.97 $\cdot 10^{-2}$ & 0.915 $\pm$ 0.093 \\
           &      & 7.40 $\cdot 10^{-2}$ & 0.884 $\pm$ 0.086 \\
           &      & 1.03 $\cdot 10^{-1}$ & 0.694 $\pm$ 0.078 \\
 \hline
 0.58-0.66 & 0.62 & 6.50 $\cdot 10^{-3}$ & 0.982 $\pm$ 0.073 \\
           &      & 2.12 $\cdot 10^{-2}$ & 0.985 $\pm$ 0.091 \\
           &      & 4.08 $\cdot 10^{-2}$ & 0.984 $\pm$ 0.089 \\
           &      & 6.68 $\cdot 10^{-2}$ & 0.694 $\pm$ 0.070 \\
           &      & 9.94 $\cdot 10^{-2}$ & 0.678 $\pm$ 0.064 \\
           &      & 1.39 $\cdot 10^{-1}$ & 0.526 $\pm$ 0.058 \\
 \hline
 0.66-0.74 & 0.70 & 8.39 $\cdot 10^{-3}$ & 0.896 $\pm$ 0.061 \\
           &      & 2.74 $\cdot 10^{-2}$ & 0.781 $\pm$ 0.071 \\
           &      & 5.27 $\cdot 10^{-2}$ & 0.726 $\pm$ 0.067 \\
           &      & 8.64 $\cdot 10^{-2}$ & 0.507 $\pm$ 0.051 \\
           &      & 1.29 $\cdot 10^{-1}$ & 0.366 $\pm$ 0.041 \\
           &      & 1.79 $\cdot 10^{-1}$ & 0.343 $\pm$ 0.041 \\
 \hline
 0.74-0.82 & 0.78 & 1.05 $\cdot 10^{-2}$ & 0.840 $\pm$ 0.053 \\
           &      & 3.43 $\cdot 10^{-2}$ & 0.664 $\pm$ 0.058 \\
           &      & 6.60 $\cdot 10^{-2}$ & 0.462 $\pm$ 0.047 \\
           &      & 1.08 $\cdot 10^{-1}$ & 0.330 $\pm$ 0.036 \\
           &      & 1.61 $\cdot 10^{-1}$ & 0.223 $\pm$ 0.028 \\
           &      & 2.24 $\cdot 10^{-1}$ & 0.162 $\pm$ 0.024 \\
 \hline
 0.82-0.90 & 0.86 & 1.28 $\cdot 10^{-2}$ & 0.364 $\pm$ 0.032 \\
           &      & 4.20 $\cdot 10^{-2}$ & 0.289 $\pm$ 0.035 \\
           &      & 8.08 $\cdot 10^{-2}$ & 0.194 $\pm$ 0.027 \\
           &      & 1.33 $\cdot 10^{-1}$ & 0.145 $\pm$ 0.021 \\
           &      & 1.97 $\cdot 10^{-1}$ & 0.044 $\pm$ 0.011 \\
           &      & 2.75 $\cdot 10^{-1}$ & 0.048 $\pm$ 0.011 \\
 \hline
 0.90-1.00 & 0.93 & 1.52 $\cdot 10^{-2}$ & 0.049 $\pm$ 0.009 \\
           &      & 5.03 $\cdot 10^{-2}$ & 0.033 $\pm$ 0.009 \\
           &      & 9.68 $\cdot 10^{-2}$ & 0.022 $\pm$ 0.007 \\
           &      & 1.59 $\cdot 10^{-1}$ & 0.006 $\pm$ 0.003 \\
           &      & 2.36 $\cdot 10^{-1}$ & 0.002 $\pm$ 0.002 \\
           &      & 3.29 $\cdot 10^{-1}$ & 0.006 $\pm$ 0.003 \\
 \hline
\end{tabular}}
\caption{
The normalized doubly differential distributions
$ \sigma_{\rm norm.} = (1/\sigma_{\inc}) d^2 \sigma_{\LN}/dx_L dp_T^2.$
Only statistical uncertainties are shown.
}
\label{tab-ptsq}
\end{center}
\end{table}

\clearpage


\begin{table}[p]
\begin{center}
\begin{tabular}{|c|c|c|c|}
\hline
$x_L$ range & $\langle x_L \rangle$ & $a \: (\gev^{-2})$ & $b \: (\gev^{-2})$  \\
\hline
 0.20--0.50 & 0.38 & $ 1.726 \pm 0.115^{+0.206}_{-0.199}$ & $  8.63 \pm 7.45^{+9.36}_{-9.52}$ \\
 0.50--0.58 & 0.54 & $ 1.084 \pm 0.072^{+0.081}_{-0.104}$ & $  4.00 \pm 1.23^{+1.24}_{-1.24}$ \\
 0.58--0.66 & 0.62 & $ 1.058 \pm 0.061^{+0.060}_{-0.035}$ & $  4.89 \pm 0.83^{+0.54}_{-0.49}$ \\
 0.66--0.74 & 0.70 & $ 0.940 \pm 0.054^{+0.027}_{-0.038}$ & $  6.46 \pm 0.72^{+0.23}_{-0.65}$ \\
 0.74--0.82 & 0.78 & $ 0.878 \pm 0.051^{+0.025}_{-0.026}$ & $  8.38 \pm 0.67^{+0.97}_{-0.72}$ \\
 0.82--0.90 & 0.86 & $ 0.420 \pm 0.033^{+0.089}_{-0.072}$ & $  9.61 \pm 0.83^{+0.71}_{-0.76}$ \\
 0.90--1.00 & 0.93 & $ 0.061 \pm 0.011^{+0.024}_{-0.026}$ & $ 12.89 \pm 2.11^{+2.66}_{-2.27}$ \\
\hline
\end{tabular}
\caption{
The intercepts $a$ and slopes $b$ of the exponential
parameterization of the differential cross section defined
in \Sect{ptsq}.
Statistical uncertainties are listed first, followed by
systematic uncertainties, not including an overall
normalization uncertainty of 2.1\% on the intercepts.
The systematic uncertainties are strongly correlated
between all points.
}
\label{tab-intslope}
\end{center}
\end{table}

\clearpage


\clearpage

\clearpage

\clearpage

\clearpage

\clearpage

\clearpage

\clearpage

\clearpage

\clearpage

\clearpage

\clearpage

\clearpage

\clearpage

\clearpage

\clearpage

\clearpage

\clearpage

%
%

\begin{mcbibliography}{10}

\bibitem{engler}
J.\ Engler \etal,
\newblock Nucl.\ Phys.{} {\bf B~84},~70~(1975)\relax
\relax
\bibitem{robinson}
B.\ Robinson \etal,
\newblock Phys.\ Rev.\ Lett.{} {\bf 34},~1475~(1975)\relax
\relax
\bibitem{flauger}
W.\ Flauger and F.\ M\"onnig,
\newblock Nucl.\ Phys.{} {\bf B~109},~347~(1976)\relax
\relax
\bibitem{hanlon}
J.\ Hanlon \etal,
\newblock Phys.\ Rev.\ Lett.{} {\bf 37},~967~(1976)\relax
\relax
\bibitem{eisenberg}
Y.\ Eisenberg \etal,
\newblock Nucl.\ Phys.{} {\bf B~135},~189~(1978)\relax
\relax
\bibitem{blobel}
V.\ Blobel \etal,
\newblock Nucl.\ Phys.{} {\bf B~135},~379~(1978)\relax
\relax
\bibitem{abramowicz}
H.\ Abramowicz \etal,
\newblock Nucl.\ Phys.{} {\bf B~166},~62~(1980)\relax
\relax
\bibitem{sullivan}
J.\ D.\ Sullivan,
\newblock Phys.\ Rev.{} {\bf D~5},~1732~(1972)\relax
\relax
\bibitem{bishari}
M.\ Bishari,
\newblock Phys.\ Lett.{} {\bf B~38},~510~(1972)\relax
\relax
\bibitem{ganguli}
S.N.~Ganguli and D.P.~Roy,
\newblock Phys.\ Rep.{} {\bf 67},~201~(1980)\relax
\relax
\bibitem{zoller}
V.\ R.\ Zoller,
\newblock Z.\ Phys.{} {\bf C~53},~443~(1992)\relax
\relax
\bibitem{pl:b384:388}
ZEUS \coll, M.~Derrick \etal,
\newblock Phys.\ Lett.{} {\bf B~384},~388~(1996)\relax
\relax
\bibitem{epj:c6:587}
H1 \coll, C.~Adloff \etal,
\newblock Eur.\ Phys.\ J.{} {\bf C~6},~587~(1999)\relax
\relax
\bibitem{np:b596:3}
ZEUS \coll, J. Breitweg \etal,
\newblock Nucl.\ Phys.{} {\bf B~596},~3~(2001)\relax
\relax
\bibitem{np:b637:3}
ZEUS \coll, S.~Chekanov \etal,
\newblock Nucl.\ Phys.{} {\bf B~637},~3~(2002)\relax
\relax
\bibitem{np:b658:3}
ZEUS \coll, S.~Chekanov \etal,
\newblock Nucl.\ Phys.{} {\bf B~658},~3~(2003)\relax
\relax
\bibitem{pl:b590:143}
ZEUS \coll, S.~Chekanov \etal,
\newblock Phys.\ Lett.{} {\bf B~590},~143~(2004)\relax
\relax
\bibitem{epj:c41:273}
H1 \coll, A.~Aktas \etal,
\newblock Eur.\ Phys.\ J.{} {\bf C~41},~273~(2005)\relax
\relax
\bibitem{pl:b610:199}
ZEUS \coll, S.~Chekanov \etal,
\newblock Phys.\ Lett.{} {\bf B~610},~199~(2005)\relax
\relax
\bibitem{DESY-07-011}
ZEUS Coll., S. Chekanov \etal,
\newblock Nucl.\ Phys.{} {\bf B~776},~1~(2007)\relax
\relax
\bibitem{pr:188:2159}
J.~Benecke \etal,
\newblock Phys. Rev.{} {\bf 188},~2159~(1969)\relax
\relax
\bibitem{pr:d50:590}
T.T.~Chou and C.N.~Yang,
\newblock Phys. Rev.{} {\bf D~50},~590~(1994)\relax
\relax
\bibitem{rescat3}
N.N. Nikolaev, J. Speth and B.G. Zakharov,
\newblock hep-ph/9708290{}~(1997)\relax
\relax
\bibitem{rescat4}
U.~D'Alesio and H.J.~Pirner,
\newblock Eur. Phys. J.{} {\bf A~7},~109~(2000)\relax
\relax
\bibitem{epj:c47:385}
A.B.~Kaidalov \etal,
\newblock Eur.\ Phys.\ J.{} {\bf C~47},~385~(2006)\relax
\relax
\bibitem{epj:c48:797}
V.A.~Khoze, A.D.~Martin and M.G.~Ryskin,
\newblock Eur.\ Phys.\ J.{} {\bf C~48},~797~(2006)\relax
\relax
\bibitem{epj:c51:549}
H1 \coll, A.~Aktas \etal,
\newblock Eur.\ Phys.\ J.{} {\bf C~51},~549~(2007)\relax
\relax
\bibitem{roger}
ZEUS Coll., S. Chekanov \etal,
\newblock Eur.\ Phys.\ J.{} {\bf C~55},~177~(2008)\relax
\relax
\bibitem{alessio}
ZEUS Coll., S. Chekanov \etal,
\newblock Eur. Phys. J.{} {\bf C~52},~813~(2007)\relax
\relax
\bibitem{zeus:1993:bluebook}
ZEUS \coll, U.~Holm~(ed.),
\newblock {\em The {ZEUS} Detector}.
\newblock Status Report (unpublished), DESY (1993),
\newblock available on
  \texttt{http://www-zeus.desy.de/bluebook/bluebook.html}\relax
\relax
\bibitem{nim:a279:290}
N.~Harnew \etal,
\newblock Nucl.\ Inst.\ Meth.{} {\bf A~279},~290~(1989)\relax
\relax
\bibitem{npps:b32:181}
B.~Foster \etal,
\newblock Nucl.\ Phys.\ Proc.\ Suppl.{} {\bf B~32},~181~(1993)\relax
\relax
\bibitem{nim:a338:254}
B.~Foster \etal,
\newblock Nucl.\ Inst.\ Meth.{} {\bf A~338},~254~(1994)\relax
\relax
\bibitem{nim:a309:77}
M.~Derrick \etal,
\newblock Nucl.\ Inst.\ Meth.{} {\bf A~309},~77~(1991)\relax
\relax
\bibitem{nim:a309:101}
A.~Andresen \etal,
\newblock Nucl.\ Inst.\ Meth.{} {\bf A~309},~101~(1991)\relax
\relax
\bibitem{nim:a321:356}
A.~Caldwell \etal,
\newblock Nucl.\ Inst.\ Meth.{} {\bf A~321},~356~(1992)\relax
\relax
\bibitem{nim:a336:23}
A.~Bernstein \etal,
\newblock Nucl.\ Inst.\ Meth.{} {\bf A~336},~23~(1993)\relax
\relax
\bibitem{nim:a450:235}
A.~Bamberger \etal,
\newblock Nucl.\ Inst.\ Meth.{} {\bf A~450},~235~(2000)\relax
\relax
\bibitem{nim:a394:121}
S.~Bhadra \etal,
\newblock Nucl.\ Inst.\ Meth.{} {\bf A~394},~121~(1997)\relax
\relax
\bibitem{desy-92-066}
J.~Andruszk\'ow \etal,
\newblock Preprint \mbox{DESY-92-066}, DESY, 1992\relax
\relax
\bibitem{acpp:b32:2025}
J.~Andruszk\'ow \etal,
\newblock Acta Phys.\ Pol.{} {\bf B~32},~2025~(2001)\relax
\relax
\bibitem{uproc:chep:1992:222}
W.H.~Smith, K.~Tokushuku and L.W.~Wiggers,
\newblock {\em Proc.\ Computing in High-Energy Physics (CHEP), Annecy, France,
  Sept. 1992}, C.~Verkerk and W.~Wojcik~(eds.), p.~222.
\newblock CERN, Geneva, Switzerland (1992).
\newblock Also in preprint \mbox{DESY 92-150B}\relax
\relax
\bibitem{epj:c1:81}
ZEUS \coll, J.~Breitweg \etal,
\newblock Eur.\ Phys.\ J.{} {\bf C~1},~81~(1998)\relax
\relax
\bibitem{thesis:briskin:1998}
G.M.~Briskin.
\newblock Ph.D.\ Thesis, Tel Aviv University, Report \mbox{DESY-THESIS
  1998-036}, 1998\relax
\relax
\bibitem{proc:epfacility:1979:391}
F.~Jacquet and A.~Blondel,
\newblock {\em Proceedings of the Study for an $ep$ Facility for {Europe}},
  U.~Amaldi~(ed.), p.~391.
\newblock Hamburg, Germany (1979).
\newblock Also in preprint \mbox{DESY 79/48}\relax
\relax
\bibitem{np:b406:187}
S.Catani \etal,
\newblock Nucl.\ Phys.{} {\bf B~406},~187~(1993)\relax
\relax
\bibitem{pr:d48:3160}
S.D.~Ellis and D.E.~Soper,
\newblock Phys.\ Rev.{} {\bf D~48},~3160~(1993)\relax
\relax
\bibitem{proc:snowmass:1990:134}
J.E.~Huth \etal,
\newblock {\em Research Directions for the Decade. Proceedings of Summer Study
  on High Energy Physics, 1990}, E.L.~Berger~(ed.), p.~134.
\newblock World Scientific (1992).
\newblock Also in preprint \mbox{FERMILAB-CONF-90-249-E}\relax
\relax
\bibitem{pl:b560:7}
ZEUS \coll, S.~Chekanov \etal,
\newblock Phys.\ Lett.{} {\bf B~560},~7~(2003)\relax
\relax
\bibitem{pythia}
T.~Sj\"{o}strand \etal,
\newblock Comp.\ Phys.\ Comm.{} {\bf 135},~238~(2001)\relax
\relax
\bibitem{cpc:67:465}
G.~Marchesini \etal,
\newblock Comp.\ Phys.\ Comm.{} {\bf 67},~465~(1992)\relax
\relax
\bibitem{prep:97:31}
B.~Andersson \etal,
\newblock Phys.\ Rep.{} {\bf 97},~31~(1983)\relax
\relax
\bibitem{cpc:82:74}
T.~Sj\"ostrand,
\newblock Comp.\ Phys.\ Comm.{} {\bf 82},~74~(1994)\relax
\relax
\bibitem{cpc:135:238}
T.~Sj\"{o}strand \etal,
\newblock Comp.\ Phys.\ Comm.{} {\bf 135},~238~(2001)\relax
\relax
\bibitem{cpc:39:347}
T.~Sj\"ostrand,
\newblock Comp.\ Phys.\ Comm.{} {\bf 39},~347~(1986)\relax
\relax
\bibitem{cpc:43:367}
T.~Sj\"ostrand and M.~Bengtsson,
\newblock Comp.\ Phys.\ Comm.{} {\bf 43},~367~(1987)\relax
\relax
\bibitem{np:b238:492}
B.R.~Webber,
\newblock Nucl.\ Phys.{} {\bf B~238},~492~(1984)\relax
\relax
\bibitem{tech:cern-dd-ee-84-1}
R.~Brun et al.,
\newblock {\em {\sc geant3}},
\newblock Technical Report CERN-DD/EE/84-1, CERN, 1987\relax
\relax
\bibitem{pl:b366:371}
A.~Edin, G.~Ingelman and J.~Rathsman,
\newblock Phys.\ Lett.{} {\bf B~366},~371~(1996)\relax
\relax
\bibitem{epj:c12:375}
CTEQ \coll, H.L.~Lai \etal,
\newblock Eur.\ Phys.\ J.{} {\bf C~12},~375~(2000)\relax
\relax
\bibitem{pr:d45:3986}
M.~Gl\"{u}ck, E.~Reya, and A.~Vogt,
\newblock Phys. Rev.{} {\bf D~45},~3986~(1992)\relax
\relax
\bibitem{zfp:c76:613}
H1 \coll, C.~Adloff \etal,
\newblock Z.\ Phys.{} {\bf C~76},~613~(1997)\relax
\relax
\bibitem{zfp:c53:651}
M.~Gl\"{u}ck, E.~Reya, and A.~Vogt,
\newblock Z.\ Phys.{} {\bf C~53},~651~(1992)\relax
\relax
\bibitem{pl:b338:363}
H.~Holtmann \etal,
\newblock Phys.\ Lett.{} {\bf B~338},~363~(1994)\relax
\relax
\bibitem{epj:c33:542}
R.~Enberg, G.~Ingelman and N.~Timneanu,
\newblock Eur.\ Phys.\ J.{} {\bf C~33},~542~(2004)\relax
\relax
\bibitem{epj:c7:609}
ZEUS \coll, J.~Breitweg \etal,
\newblock Eur.\ Phys.\ J.{} {\bf C~7},~609~(1999)\relax
\relax
\bibitem{np:b713:3}
ZEUS \coll, S.~Chekanov \etal,
\newblock Nucl.\ Phys.{} {\bf B~713},~3~(2005)\relax
\relax
\end{mcbibliography}
\end{document}